# Reinforced room temperature spin filtering in chiral paramagnetic metallopeptides

Ramón Torres-Cavanillas[1], Garin Escorcia-Ariza[1], Isaac Brotons-Alcázar[1], Roger Sanchís-Gual[1], Prakash Chandra Mondal[1], Lorena E. Rosaleny[1], Silvia Giménez-Santamarina[1], Michele Sessolo[1], Marta Galbiati[1], Sergio Tatay[1], Alejandro Gaita-Ariño[1*], Alicia Forment-Aliaga[1*], Salvador Cardona-Serra[1*]

[1]ICMol. Universitat de València. C/ Catedrático José Beltrán nº 2, 46980 Paterna, Valencia, España.

**ABSTRACT:** Chiral-induced spin selectivity (CISS), whereby helical molecules polarize the spin of electrical current, is an intriguing effect with potential applications in nanospintronics. In this nascent field, the study of paramagnetic chiral molecules, which could introduce another degree of freedom in the control of the spin transport, remains so far unexplored. To address this challenge, herein, we propose the use of self-assembled monolayers of helical lanthanide-binding peptides. In order to elucidate the effect of the paramagnetic nuclei, monolayers of the peptide coordinating paramagnetic or diamagnetic ions are prepared. By means of spin-dependent electrochemistry, CISS effect is demonstrated by cyclic voltammetry and impedance measurements for both samples. Additionally, an implementation of the standard liquid-metal drop electron transport setup has been carried out, demonstrating their suitability for solid-state devices. Remarkably, the inclusion of a paramagnetic center in the peptide increases the spin polarization as independently proved by different techniques. These findings permit the inclusion of magnetic biomolecules in the CISS field, paving the way to their implementation in a new generation of spintronic nanodevices.

The discovery of the giant and tunnel magnetoresistance gave rise to the field of spintronics:[1,2] the use of the electronic spin in addition to the electronic charge to transport and process information. Typically, inorganic materials had been used as active elements of such devices. However, the study of spin-related phenomena in molecular systems have garnered considerable attention in the last years, allowing the extension from inorganic into molecular spintronics. Since then, researchers have started to profit from the rich diversity of molecules that chemistry and nature can provide, to combine and improve the functional components of a spintronic device. Molecular Spintronics is young but prolific, and recent experimental advances [3-5] have been supported by the theoretical evaluation of various molecular families [6,7,] and DNA strands[8].

In this scenario, chiral molecules are playing a singular role in the field since Naaman and others showed that the electronic transport along chiral diamagnetic molecules produces a net spin polarization of the current, the so-called Chiral Induced Spin Selectivity (CISS)[9], and thus can behave as a molecular-based spin filter. The origin of this phenomenon has been deeply studied by theoretical calculations, occasionally combined with experiments [10-16] and it has been recently proposed to be of relevance in different issues like the role of spin in electron transfer through bio-systems[17], the building of spintronic logic devices based on organic ferromagnets at room temperature[18] or the fabrication of a spintronic magnetic memory which, including a ferromagnetic platelet, presents memristor-like behavior[19]. Since its discovery, the CISS effect has been studied, both in single molecules[20] and in thin layers of a broad spectrum of chiral molecular systems, ranging from Langmuir-Blodgett films of L- or D- stearoyl lysine[21] to self-assembled monolayers (SAMs) of polyalanine [22], double-strand DNA [23], bacteriorhodopsin[24,25] and helicene molecules[26]. On the other hand, the possibility to induce room temperature spin-dependent transport has also been probed through molecules with a single paramagnetic ion[27], and the specific electronic configuration of the metal centers has turned out to be decisive for the spin-polarized current [28]. However, so far, the combination of chirality and paramagnetism in molecular spin filters has remained elusive.

In this context, we identified the family of helical metallopeptides known as Lanthanide Binding Tags (LBTs) as a promising system for CISS-based biospintronics. LBTs were originally designed by Imperiali *et al.* to achieve high affinity for lanthanoid ions,[29-32] and have been known to autonomously and robustly fold into a partially helical structure ($3_{10}$ helix), where one of the atoms in the coordination sphere directly coordinates the lanthanoid ion ($Ln^{3+}$) through a carbonyl group belonging to a peptide bond [33]. Since the current in peptides has been known to flow through the backbone chain[34], some reciprocal influence between the spin polarization of the current along the helical backbone and the spin orientation of the lanthanoid ion could be expected.

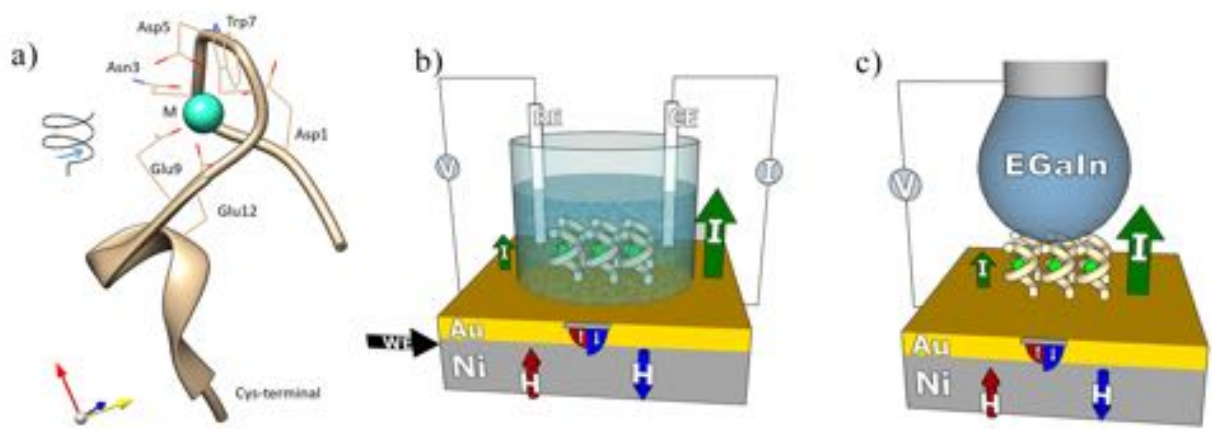

**Fig. 1** LBTC metallopeptide and spin filtering. **a** Side view of MLBTC in ribbon representation (based on a structure with PDB ID:1TJB); arrows show the right-handed helicity. The aminoacids coordinating the metal ion are tagged (Asp1, Asn3, Asp5, Trp7, Glu12, Glu14), including those coordinating through different lateral residues and the Trp containing the coordinating carbonyl oxygen. **b** and **c** Schemes of spin-dependent transport setups used in this work emphasizing the magnetization of the Ni layer under an external magnetic field (350 mT) and its influence on the empty and occupied density of states, as well as the expected CISS effect of the right-handed helical MLBTC SAM. **b** Three-electrode electrochemical cell: MLBTC SAM on (Ni/Au) as working electrode (WE), Ag/AgCl reference electrode (RE) and Pt wire as a counter electrode (CE) in a buffered 5 mM $K_4Fe(CN)_6/K_3Fe(CN)_6$ solution as an external redox probe. **c** Home-made liquid-metal EGaIn transport system.

Recently, our group has shown that they can be used as spin tags to open a fruitful playground in the field of molecular spin qubits [35], thanks to the ease of preparing new structures with optimized properties by means of standard techniques of molecular biology. Furthermore, LBTs have also been used to obtain insight into the real-time interaction between the dynamics of spin energy levels and internal molecular motions [36]. Here, we prove the spin polarization effect in LBT SAMs and demonstrate a non-innocent role of the paramagnetic center. By means of spin dependent electrochemistry [37], spin dependent local transport in solid state devices using liquid-metal drop contacts, and DFT transport calculations we show the potential of LBT SAMs for spin filtering beyond the CISS effect.

### Results and discussion

**Surface assembling of metalloLBTC units**

For our experiments, we selected a LBT sequence [33] modified with an additional cysteine (LBTC, see Fig. 1a) as the C-terminal position to ensure a strong Metal-S interaction between the biomolecule and a metallic surface like Au or Ni which is an essential requirement for the performance of spin-dependent transport measurements. We prepared the peptide and metallopeptide-based SAMs by overnight immersion in aqueous buffered LBTC or LBTC with $Tb^{3+}$ (TbLBTC) solutions and a reducing agent (tris(2-carboxyethyl)phosphine) to avoid thiol oxidation during the grafting process (see SI section S1). Anchoring of the molecules on the surface was confirmed by the identification of the LBTC molecular weight in the fragments extracted in time-of-flight mass spectrometry (see SI section S3), and by the chemical composition of the functionalized surfaces with X-ray photoelectron spectroscopy which shows the presence of terbium (see SI section S4). Moreover, quartz crystal microbalance measurements suggested near complete molecular coverage (see SI section S6) and atomic force microscopy (see SI section S2) gave us insight into their homogeneous topography. The persistence of the lanthanide ion bound to the LBTC peptide on the surface was experimentally supported by its luminescence emission (see SI section S1, Fig. S1.2). The stability of the three-dimensional folding is consistent with our estimation of the SAM thickness (see SI section S7). More importantly, previous works have demonstrated by using the SHAPE program [35] and Q values [38] the high similarity between different reported crystal structures as well as solution structures of several LnLBT metallopeptides. [38] It is well established now that the folding of LBT peptides is very robust and essentially indistinguishable in solution or in solid state. This means no relevant change in helicity or metal coordination is expected in our SAMs regardless of the coordination atom. Instead, a narrow distribution of shapes is expected for the peptides at room temperature due to thermal effects. [36]

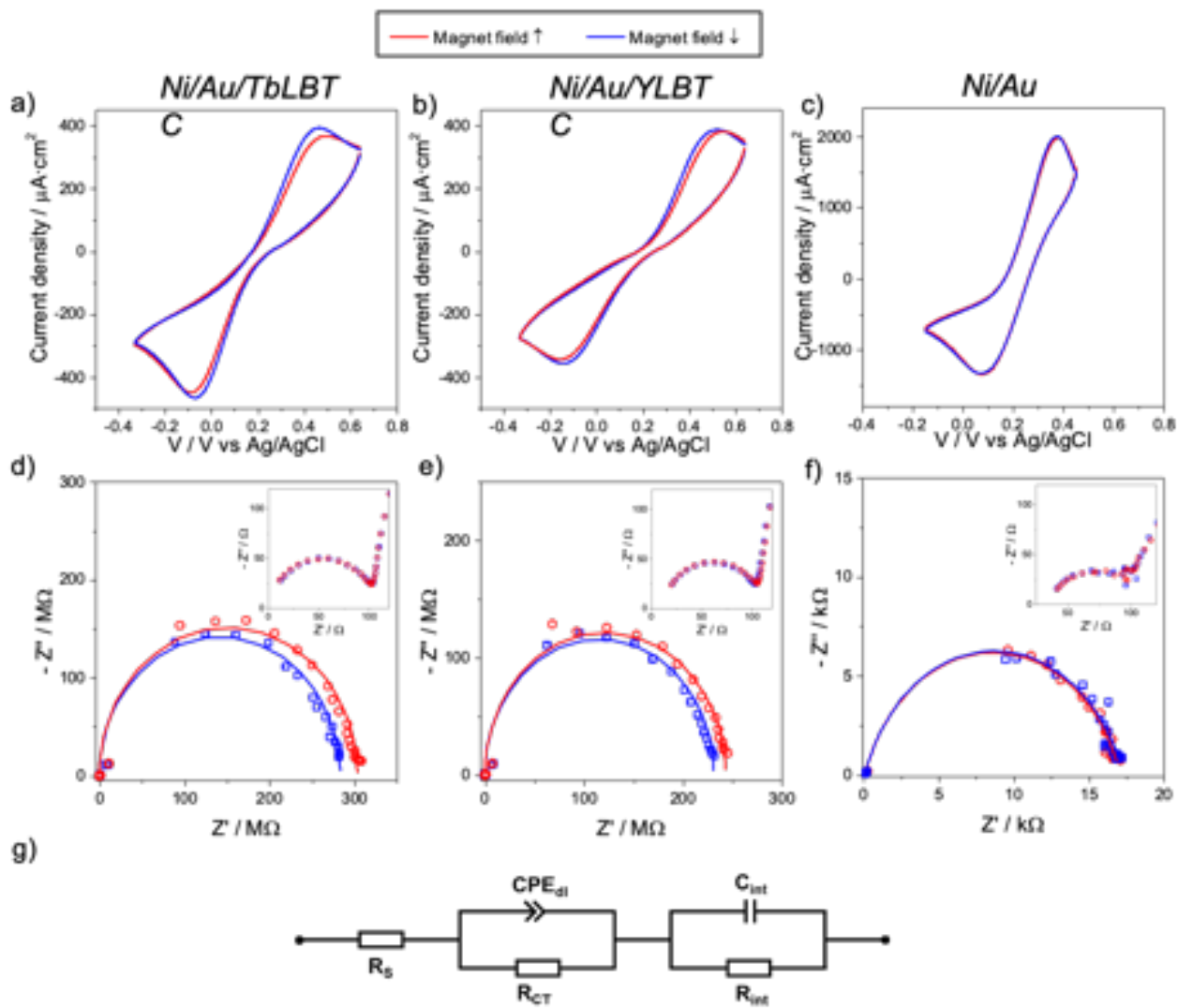

**Fig. 2** Magnetic field-dependent electrochemistry. **a-c** Spin-dependent cyclic voltammograms on a SAM of TbLBTC on Ni/Au (a) and YLBTC on Ni/Au (b), and on the bare Ni/Au working electrode (c). Voltammograms were recorded at 20 mV/s under an external magnetic field pointing "up" (solid red line) or "down" (solid blue line); 4 cycles were averaged. **d-f** Electrochemical impedance spectroscopy (EIS) of a SAM of TbLBTC on Ni/Au (d) and YLBTC on Ni/Au (e) and on a bare Ni/Au working electrode (f), under an external magnetic field pointing "up" (red circles) or "down" (blue squares). **e** The equivalent circuit employed for the EIS fit.

**Spin-dependent electrochemistry**

Once the possibility of assembling metalloLBTC (MLBTC) complexes onto metallic surfaces was demonstrated, their spin filtering properties were investigated by spin-dependent electrochemical measurements. For these experiments, $K_4Fe(CN)_6/K_3Fe(CN)_6$ was used as a redox active external probe due to the non-electroactive nature of the SAM. As expected, scan rate dependence measurements confirmed that this redox reaction is limited by a diffusion process [39] (see SI section S8). Next, the integration of a permanent magnet in the electrochemical setup permitted to perform spin-dependent cyclic voltammetry, which is a well-established technique that is useful to measure the spin-selective electron conduction (as shown in Fig. 1b) [37] between the measured $K_4Fe(CN)_6/K_3Fe(CN)_6$ voltammograms when MLBTC-modified (Fig. 2a-b) or a bare ferromagnetic Ni/Au electrodes (200 nm Ni protected with 5 nm Au) were used (Fig. 2c). The results showed that the current in the case of the bare Ni/Au working electrode (WE) was quasi-independent of the orientation of the magnet placed underneath, with the differences for each magnetic field direction being comparable with those observed for two consecutive voltammetry scans (Fig. 2c). However, when the MLBTC-modified WE was used, the current density was appreciably higher with the external magnet pointing "down" compared to when it was pointing "up". Furthermore, a decrease in the current and an anodic and catodic shift in the oxidation and reduction potential of the Ni/Au/MLBTC system were observed (Fig. 2a-b). This kind of changes in the voltammetric cycles are expected due to the presence of a less conductive layer between the electrochemical probe and the metal in the WE.

We define the spin polarization (*SP*) as:

$$SP = \frac{J_{up} - J_{down}}{J_{up} + J_{down}} \times 100$$

, where $J_{up}$, and $J_{down}$ stand for current density ($J_i = I_i/Area$) at specific potential when the Ni layer is magnetized "up", and "down", respectively. For the Ni/ Au/ TbLBTC, $SP$ = -3.3 (± 0.2)% measured at +460 mV, and -1.9 (± 0.2)% at -70 mV (vs. Ag/AgCl) with scan rate of 20 mV·s$^{-1}$ were obtained. Moreover, the anodic and catodic peak shifts -28 mV and -20 mV, respectively. The negative sign means that electrons are preferentially transported with their spin orientation antiparallel to the direction of their movement[40].

These results are consistent with the expected effect of the right-handed helicity of the molecule in our experimental setup (Fig 1b)[40]. The external Nd magnet below the TbLBTC modified substrate governs the orientation of the magnetization of the Ni layer, which in turn determines the energy of the spin-up vs. the spin-down bands in the Au layer near the Fermi level. When the external magnet is pointing up, the Ni valence band will be polarized spin up, and the Ni conduction band will be polarized spin down, and vice-versa. Independently of the magnet direction, right-handed helices always tend to spin-polarize moving electrons (current) in the direction antiparallel to the direction of their movement. For current flowing upwards through our right-handed SAM, this means the majority current will be spin down; here we represent this as a larger spin-down density of states at the Au-SAM interface. For our experiments, this implies that, with the magnet pointing down, the system is expected to conduct better, whether this means spin down electrons flowing upwards (extracted from the Ni valence band) or spin up electrons flowing downwards (injected into the Ni conduction band).

Interestingly, when the diamagnetic $Y^{3+}$ center replaces the paramagnetic $Tb^{3+}$, SP effect was slightly decreased, displaying $SP$ of -0.6 (± 0.2)% measured at $E_{pc}$ = +515 mV, and -1.9 (± 0.2)% at $E_{pa}$ = -140 mV (vs. Ag/AgCl). Additionally, the anodic and catodic shift of the oxidation and reduction potential also decreases respect its paramagnetic counterpart, -21 and -11 mV respectively (Fig 2a-b).

This experimental observation suggested an increase in the filtering of the SAM, product of the magnetic behaviour of the metallic center. Nevertheless, due to the high voltages applied to the sample during the measurements, modification in the SAM packing can occur, complicating the quantitative comparison of the SP between samples.

Therefore, to confirm the reliability of this experimental observation, Electrochemical impedance spectroscopy (EIS) measurements were carried out. EIS is a powerful technique to further investigate and characterize electrochemical processes. While typical direct current voltammetry requires the application of high voltage, up to 0.5 V, potentially affecting the packing of the SAM, lower voltages are applied in EIS (voltage amplitude of 0.01V). Therefore, the structural rearrangements produced by the flow of current are minimized.

For the evaluation of electron-transfer properties under the influence of an external magnetic field pointing either "up" or "down", an AC potential of 10 mV going from $10^5$ to $10^0$ Hz was applied at the Open Circuit Potential. Fig. 2d and e displays the Nyquist plot of the response of the corresponding electrochemical systems. The equivalent circuit used to fit the data is shown in Fig. 2g. Constant phase element (CPE), which is a non-ideal capacitance, is introduced to provide a good match with the experimental data because of the possible surface roughness, physical non-uniformity and/or the non-uniform distribution of the electroactive sites. The equivalent circuit is composed by a resistance coming from the ionic transport through the solution ($R_s$) connected in series with a first parallel branch ($R_{int}$ and $C_{int}$) corresponding to the electron that has to move through the current collectors and the different layers of the WE. These three elements appear in the high-frequency region. On the other hand, the low-frequency region is associated with a second parallel branch corresponding to the charge transfer and its interface capacitance ($R_{CT}$ and $CPE_{dl}$) taking place between the SAM and the electrolyte. As it can be noticed, at high frequencies, the first semicircles remain unaltered after modifying the magnetic field sense. Contrary to that, the second semicircles, that involve the redox process of $K_4Fe(CN)_6/K_3Fe(CN)_6$, are clearly affected. With the external magnet pointing "down", these semicircles are reduced. In this direction, the more depressed the semicircle in the Nyquist plot, the lower the charge transfer resistance. Thus, a reduction of the resistance is observed by changing an external magnet from "up" to "down". A similar trend was observed starting in "down" position and changing to "up" (see Fig. S8.3).

Remarkably, it was found a $SP$ of -3.6 (± 0.2) % for the TbLBTC and -2.4 (± 0.2) % for the YLBTC. This strongly points out to an active role of the metallic center in the spin selectivity. A blank measurement (i.e. without peptide and metal) was also carried out to ascertain the spin polarization effect. In Fig. 2f it can be seen that the peptide absence facilitates the charge transfer leading to an important decrease of $R_{CT}$ (around 4 orders of magnitude). Importantly, both semicircles exhibit almost no variation after altering the magnetic field position. Indeed, these small differences are reliably related to experimental error.

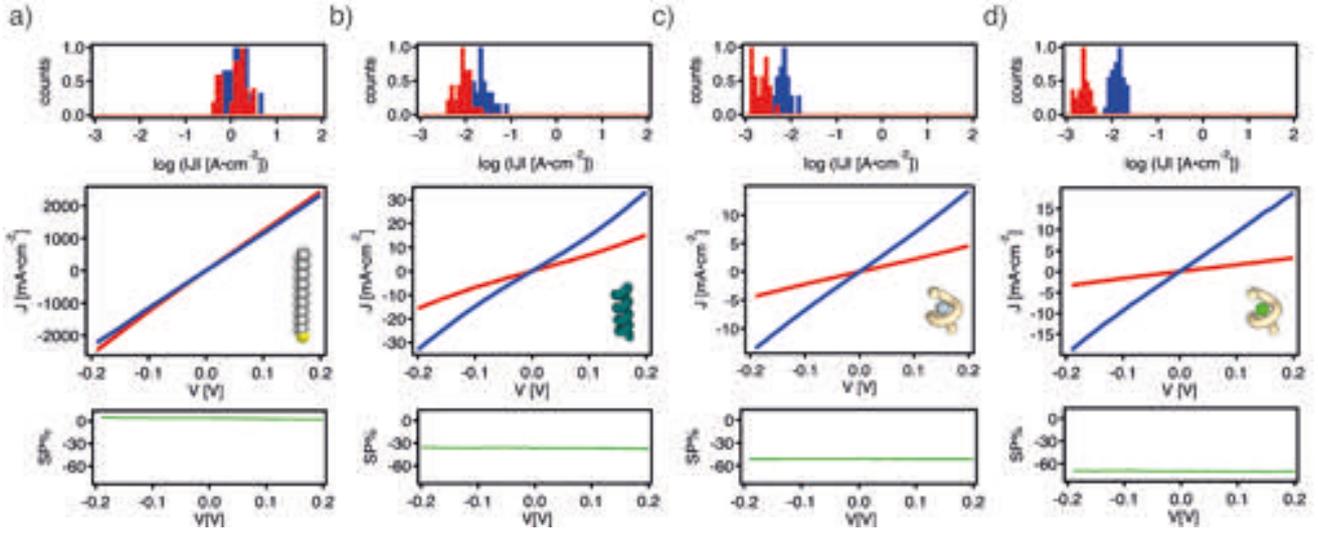

**Fig. 3** Magnetic field-dependent liquid-metal transport experiments measured for **a** C18S **b** Ala8 **c** YLBTC **d** TbLBTC. Current density histograms measured at ±0.15V (top) and averaged current density vs. voltage curves (middle) measured under an external magnetic field pointing "up" (red) or "down" (blue). *SP* polarization as a function of the bias voltage calculated from the measured averaged current density (bottom). In all cases, the sign criterion is such that negative *SP* means the device is more conductive with the magnetic field pointing down

**Liquid metal EGaIn transport measurements**

Once the spin selectivity was demonstrated, to advance towards a setup that is closer to a practical solid-state device permitting direct local spin-transport measurements, we investigated magnetic field dependent charge-transport through MLBTC with liquid-metal contacts using the Gallium Indium Eutectic alloy (EGaIn) as liquid-metal contacts (Fig. 1c). Liquid-metal drop local transport in solid state devices has been widely used in electronic current transport measurements through organic layers[41], and here we have extended it to CISS measurements by including an additional local magnetic field.
Before performing any spin-dependent measurement, to gain local information on the electron transport through MLBTC SAMs, we performed charge-transport studies using Au/TbBTC/Ga$_2$O$_3$/EGaIn liquid contact junctions. For statistics, a total of 30 different junctions with areas ranging from ~3200 to ~12000 μm$^2$ were measured with about 16 *I-V* scans (between −0.5 and 0.5 V) collected on each one. Compared to the pristine samples (without SAM), in the case of the TbBTC SAM the vast majority of the junctions (>90%) showed lower current densities with log (|*J*/A·cm$^{-2}$|) = ~ -1, roughly three orders of magnitude lower than the pristine sample that displayed log (|*J*/A·cm$^{-2}$|) = ~ 2, both measured at 0.15V (Fig. S9.1a).

To estimate the energy offset ($\varepsilon_o$) between $E_F$ (Fermi energy) and the energy of the nearest conduction orbital, $\varepsilon_o = |E - E_F|$, we used a single-level model with Lorentzian transmission (Newns–Anderson model)[42]. Fitting of the *I–V* curves up to ±500 mV yielded $\varepsilon_o$ = 0.68 ± 0.18 eV on average (a representative *I–V* curve and its fit is shown in Fig. S9.1b). This value is close to that measured using transition voltage spectroscopy (TVS). In order to measure the transition voltage ($V_t$), we collected additional *I-V* scans up to ±1 V. A Fowler-Nordheim plot reveals an average value $V_t$ = 0.82 ± 0.10 V and $\varepsilon_o$ = 0.71 eV (SI Fig. S9.1c). This information combined with the change in the absolute value of the electrode work function (Φ) due to the adsorbed TbLBTC SAM, $\Phi_{SAM}$ = 4.62 eV (measured through a combination of Kelvin probe and ambient pressure photoemission spectroscopy, see SI section S10), led to the energy level alignment scheme of Au/TbLBTC//Ga$_2$O$_3$/EGaIn junction proposed in Fig. S9.2.

In order to exploit the spin filtering effect in a solid state device, we extended the experimental setup with the addition of a permanent magnet (350 mT) in close proximity to the sample (with an additional ferromagnetic Ni layer). To validate our experimental setup, we prepared and measured two analogous devices based on SAMs of different molecules, including chiral and non-chiral SAMs. The non-chiral molecule was octadecanethiol (C18S) and the chiral reference was the right-handed HS−CH$_2$−CH$_2$−NHCO(Ala-AiB)$_8$−NH$_2$ (Ala8) [43], where Ala = Alanine and AiB = 2- aminoisobutyric acid. We measured from 6 to 22 consecutive *I-V* curves for a given drop size. From these data, we calculated the current density *J* vs. applied voltage *V*. Once up to 6 non short-circuited *I-V* curves were collected, the sample was displaced laterally with an XYZ-stage, where a new EGaIn drop was produced, and the characterization was repeated all over again. To improve the significance level, we repeated this process at least 10 times on randomly selected different parts of the film surface and collected at least 60 *I-V* curves for each sample. From these data, we calculated the current density *J* vs. applied voltage *V* for magnet "up" and "down" orientations. Fig. 3 shows current density histograms measured at ±0.2 V for opposing directions of the external magnet.

In the case of the non-chiral C18S SAM, no difference in the current density histograms for opposing directions of the external magnet were observed (Fig. 3a), in contrast to Ala8 case (Fig. 3b). Therefore, negative *SP* was only observed for helical molecules (*SP* ≈ 2 ± 2% for C18S, *SP* ≈ -40 ± 15 % for Ala8).

Once the feasibility of the set up was proved, solid state measurements of the TbLBTC and YLBTC were carried out. It must be noticed that, similarly to Ala8 SAMs, MLBTC SAMs chirality is expected to be right-handed, and thus to provide a negative *SP*.[40] As was already anticipated by the electrochemistry measurements, the *SP* differs according the metallic center present in the SAM. In both cases negative *SP* are recorded, however the filtering effect on the spin selectivity displayed by the paramagnetic SAM is considerably larger than the one displayed by its diamagnetic analogue (*SP* ≈ -50 ± 20 % for YLBTC and *SP* ≈ -70 ± 10 % for TbLBTC, Fig. 3). Noticeably, *SP* measured liquid metal contacts are of the same sign but one order of magnitude larger compared with those obtained by electrochemistry methods. We attribute this fact to the more local nature of this technique. The unavoidable presence of pin-holes in the SAM should result in a supply of unpolarized electrons that will contribute to the total measured current, and in turn this will account for a reduction of the measured *SP*. In electrochemical measurements, due to the larger contact surface, which is about 1500 times the contact area generated with EGaIn drops, this effect is maximized [44]. While, in the case of electrochemical measurements the presence of TbLBTC SAMs results in a moderate decrease of *J* compared to the pristine substrate, a three order of magnitude decrease is measured in the case of liquid metals.

**Influence of the paramagnetic $Ln^{3+}$ on transport properties.**

We now aim at estimating the upper limit of the spin filtering effect created by a lanthanide ion (which is here assumed to be fully polarized) on the current, initially neglecting the effect of the CISS. As a first step, we carried out structural relaxation calculations using the software SIESTA [45]. After this step, we employed a non-equilibrium Green function approach, using the code SMEAGOL[46] to obtain the transport properties of the junction (see SI section S11 for details). We used the calculated Density of States (DOS) to obtain the final spin-dependent transmission spectra normalized to the Fermi energy ($E_F$), as shown in Fig. 4b. We found a small transmission peak at $E-E_F$ = -0.3 eV, which, when integrating between -0.5 eV and +0.5 eV, results in a spin filtering of about *SP* = -70% assuming a complete polarization of the lanthanide ion, i.e. at very high magnetic fields or very low temperatures. The position of the peak is at a smaller voltage compared with the experimentally estimated $|E-E_F|$ = 0.7 eV but still comparable with the experimentally proposed energy level alignment scheme (Fig. S9.2).

We then studied the Local Density of States at $E-E_F$ = -0.3 eV, and confirmed that the nearest transmission orbital is indeed extended all over the peptide backbone (see Fig. S11.2). It means that this path strongly overlaps with the *f* magnetic orbitals of the lanthanide ion. In other words, the magnetic orbitals of the lanthanoid participate in the singly occupied molecular orbital (SOMO) transmission peak that appears at moderate voltages. Due to this uncommon feature, the effective exchange coupling is expected to be anomalously large. In fact, it is deemed to be within the typical range for lanthanide ions directly coordinated to organic radicals, that corresponds to a magnetic exchange coupling $J_{ex}$ = 1–25cm$^{-1}$ [47-50]. In this particular case, we estimated this level splitting with additional DFT calculations, (see SI section S12) obtaining differences between the ferro- and antiferromagnetic states of the order of ~5 cm$^{-1}$.

This calculated upper limit of *SP* = 70% to the effect of the Tb is an obvious overestimation in this case, considering it would require the $Tb^{3+}$ ion to be fully polarized. In the opposite extreme, the effect derived from a magnetic polarization in thermal equilibrium is estimated to have a negligible effect (*SP* = 51%, see details in SI section S12). However, this is expected to be an underestimation. We need to recall that, as long as current is circulating, the system is out of equilibrium. In particular, since this current is spin polarized by the CISS effect, the magnetization of the $Tb^{3+}$ ion will also be out of equilibrium. The actual result will thus lie between the two extreme scenarios, coinciding with the experimental observation of a modest, but evident, effect of the lanthanoid ion reinforcing the spin polarization, which matches our experimental observation.

**Conclusions**

Spin filtering enhancement induced by the encapsulated paramagnetic ion has been demonstrated in lanthanide binding peptides. The active role of the paramagnetic ion has been unambiguously confirmed by three independent experimental approaches: cyclic voltammetry, electrochemical impedance spectroscopy and local transport in solid state devices using liquid-metal drop contacts. In the best conditions, a spin polarization *SP* = -70 ± 10 % was achieved with the contribution of complexed $Tb^{3+}$ ions.

The interaction between paramagnetism and chirality in the same molecule is intriguing, since they are two fundamentally different mechanisms for spin filtering. Spin-dependent transport calculations have been performed to shed some light into the experimentally observed behavior, and confirmed an interaction between the magnetic polarizations of the current and the lanthanide ion.

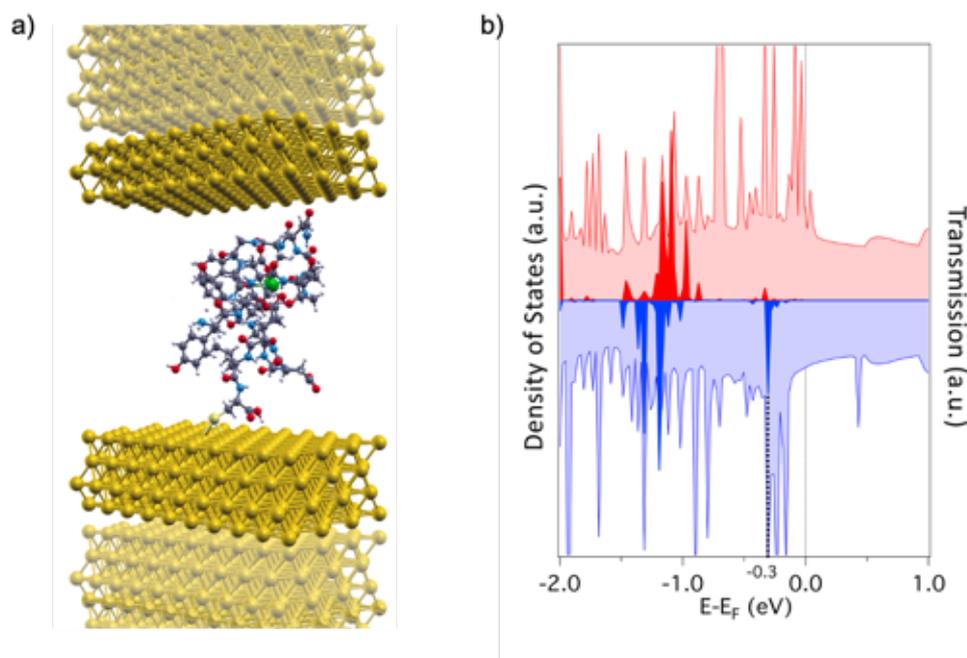

**Fig. 4** Calculated spin filtering effect due to the paramagnetic lanthanide **a** Structure of the total region (including the electrodes) that has been used as model for the calculation. Note that the system has periodic boundary conditions in *x-y* metal plane while no boundary is considered in the transport *z* direction where the bulk Au electronic structure is simulated by the non equilibrium Green's function method (depicted as transparent gold layers in the scheme). **b** Calculated Density of States (light colors) and normalized transmission spectra (strong colors) at zero voltage, distinguishing between the transmission for spin-up (red) and spin-down (blue), where the up-down reference is the polarization direction of the *f* electrons in the lanthanide. A sharp spin-down conduction peak near the Fermi level is marked with a dashed line as the main conducting peak at low gate voltages.

## Methods

**SAM preparation.** The metallopeptide-based SAMs are prepared following a similar strategy from the typical growth in solution method used for alkanethiol on gold. Substrates made of thermally evaporated metal on a $Si/SiO_2$ wafer are immersed during 24 h in a 0.1 mM buffered solution of LBTC in the presence of an excess of $Tb^{3+}$. This procedure ensures formation of a monolayer with high surface coverage as observed experimentally (Supplementary Information sections S1-S5 including luminescence studies, AFM, MALDI-TOF, XPS, and QCM).

**Spin-dependent cyclic voltammograms and Electrochemical impedance spectroscopy.** LBTC grafted on 5 nm Au overlayered onto a 200 nm Ni (Ni/Au) is used as working electrode, Pt, and Ag/AgCl are used as counter and reference electrodes, respectively. Freshly prepared and deoxygenated 5 mM $Fe^{2+}/Fe^{3+}$ solution in HEPES buffer are used as the redox mediator, and oxidation, reduction currents were monitored under an external magnetic field of 0.35 T applied underneath the TbLBTC modified ferromagnetic Ni. The same set up was used for EIS, but using Gamry 1000E and Gamry 5000E potentiostat/galvanostat controlled by Gamry software. An AC voltage of 10 mV in the frequency range of $10^0$–$10^5$ Hz at the OCP was applied. EIS data were analyzed and fitted by means of Gamry Echem Analyst v. 7.07 software.

**EGaIn measurements.** Spin-dependence by EGaIn method. LBTC grafted on 5 nm Au overlayered onto a 200 nm Ni (Ni/Au) is used as the bottom electrode and EGaIn drop as the top contact. 6 junctions were measured with 16 I-V curves between −0.2 and 0.2 V for a magnetic field applied of 0.35 T and -0.35 T by a permanent magnet placed under the metallic surface (see details in S6).

**Electron transport calculations.** First-principles calculations are performed using the SMEAGOL code that interfaces the non-equilibrium Green's function approach (NEGF) to electron transport with density functional theory (DFT). In our simulations, the transport junction is constructed by placing the polypeptide between two Au (111)-oriented surfaces. The exchange-correlation potential is described by the GGA (PBE) functional. The Au-valence electrons are represented over a numerical s-only single-ζ basis set that has been previously demonstrated to offer a good description of the energy region around the Fermi level. In contrast, for the other atoms we use a full-valence double-ζ basis set. Finally, the spin-dependent current flowing through the junction is calculated from Landauer-Büttiker formula.

**Data availability.** The data that support the findings of this study are available from the corresponding author on request.

**Author Contributions.**
R.T.-C., A.F.-A. and L.R. prepared and characterized the SAMs. M.S. and M. G. did the spectroscopic characterization of the SAMs. S.G.-S. and S. C.-S. performed the theoretical calculations. S.T. and G. E.-A. performed the EGaIn experiments, P.C.M., I.B.-A. and R.S.-G. performed the electrochemical studies. S.C.-S., A.G.-A. and A.F.-A. conceived the work and coordinated the writing of the paper.

## Acknowledgements

The research reported here was supported by the Spanish MINECO (Grants MAT2017-89528 and CTQ2017-89993, and Excellence Unit María de Maeztu MDM-2015-0538), the European Union (ERC-CoG DECRESIM 647301, FATMOLS 862893 and COST-MOLSPIN-CA15128 Molecular Spintronics Project), the Generalitat Valenciana (Prometeo Program of Excellence) and the Universitat de València (PRECOMP14-202646). A.G.-A. and M. S. thank the Spanish MINECO for the Ramón y Cajal Fellowship. S. T. thanks the Spanish MINECO for the Ramón y Cajal Fellowship (RYC-2016-19817). S.C.-S. and M. G. thank the Spanish MINECO for a 'Juan de la Cierva-Incorporación' grant. P.C.M. thanks European Union for Marie-Curie Post-doctoral Fellowship (H2020-MSCA-2015-706238). We thank Fernando Coloma (SSTTI Universidad de Alicante) for the XPS measurements and Angel López for the ellipsometry measurements (ICMol – UVEG). The MALDI-TOF analysis was carried out in the SCSIE Universitat de València Proteomics Unit, a member of ISCIII Proteo-RedProteomics Platform by O. Antúnez and L. Valero.

## REFERENCES

1 Baibich, M. N.; Broto, J. M.; Fert, A.; Van Dau, F. N.; Petroff, F.; Etienne, P.; Creuzet, G.; Friederich, A.; Chazelas, J. Giant Magnetoresistance Of (001) Fe/(001) Cr Magnetic Superlattices. *Phys. Rev. Lett.* **1988**, *61*, 2472.
2 Binasch, G.; Grünberg, P.; Saurenbach, F.; Zinn, W. Enhanced Magnetoresistance in Layered Magnetic Structures with Antiferromagnetic Interlayer Exchange. *Physical Review B* **1989**, *39*, 4828.
3 Barraud, C.; Seneor, P.; Mattana, R.; Fusil, S.; Bouzehouane, K.; Deranlot, C.; Graziosi, P.; Hueso, L.; Bergenti, I.; Dediu, V.; Petroff, F.; Fert, A. Unravelling the role of the interface for spin injection into organic semiconductors *Nat. Phys.* **2010**, *6* (8), 615–620.
4 Bogani, L.; Wernsdorfer, W. Molecular spintronics using single-molecule magnets *Nat. Mater.* **2008**, *7* (3), 179–186.
5 Forment-Aliaga, A.; Coronado, E. Hybrid Interfaces in Molecular Spintronics. *Chem. Rec.* **2018**, *18*, 737-748.
6 Cardona-Serra, S.; Gaita-Ariño, A.; Stamenova, M.; Sanvito. S. Theoretical Evaluation of [VIV(α-C3S5)3]2– As Nuclear-Spin-Sensitive Single-Molecule Spin Transistor. *J. Phys. Chem. Lett.* **2017**, 3056.
7 Cremades, E.; Pemmaraju, C. D.; Sanvito, S.; Ruiz, E. Spin-Polarized Transport Through Single-Molecule Magnet $Mn_6$ Complexes. *Nanoscale.* **2013**, *5*, 4751.
8 Pemmaraju, C. D.; Rungger, I.; Chen, X.; Rocha, A. R.; Sanvito, S. Ab Initio Study Of Electron Transport In Dry Poly (G)-Poly (C) A-DNA Strands. *Phys Rev B* **2010**, *82*, 183.
9 Gohler, B.; Hamelbeck, V.; Markus, T. Z.; Kettner, M.; Hanne, G. F.; Vager, Z.; Naaman, R.; Zacharias, H. Spin Selectivity In Electron Transmission Through Self-Assembled Monolayers Of Double-Stranded DNA. *Science.* **2011**, *331*, 894.
10 Gutierrez, R.; Díaz, E.; Naaman, R.; Cuniverti, G. Spin-selective transport through helical molecular systems. *Phys. Rev. B*, **2012**, *85*, 081404.
11 Guo, A.-M., Sun, Q. F. Spin-selective Transport of Electrons in DNA Double Helix. *Phys. Rev. Lett.* **2012**, *108*, 218102.
12 Kettner, M.; Maslyuk, V. V.; Nürenberg, D.; Seibel, J.; Gutierrez, R.; Cuniberti, G.; Ernst, K.-H.; Zacharias, H. Chirality-Dependent Electron Spin Filtering by Molecular Monolayers of Helicenes. *J. Phys. Chem. Lett.* **2018**, *9*, 2025-2030.
13 Nürenberg, D.; Zacharias, H. Evaluation of Spin-Flip Scattering in Chirality-Induced Spin Selectivity Using the Riccati Equation. *Phys. Chem. Chem. Phys.* **2019**, *21* (7), 3761–3770.
14 Medina, E.; González-Arraga, L. A.; Finkelstein-Shapiro, D.; Berche, B.; Mujica, V. Continuum Model for Chiral Induced Spin Selectivity in Helical Molecules. *J. Chem. Phys.* **2015**, *142* (19).
15 Yeganeh, S.; Ratner, M. A.; Medina, E.; Mujica, V. Chiral Electron Transport: Scattering Through Helical Potentials. *J. Chem. Phys.* **2009**, *131* (1)
16 Maslyuk, V. V.; Gutierrez, R.; Dianat, A.; Mujica, V.; Cuniberti, G. Enhanced Magnetorresistance in Chiral Molecular Junctions. *J. Phys. Chem. Lett.* **2018**, *9*, 5453-5459.
17 Eckshtain-Levi, M.; Capua, E.; Refaely-Abramson, S.; Sarkar, S.; Gavrilov, Y.; Mathew, S. P.; Paltiel, Y.; Levy, Y.; Kronik, L.; Naaman, R. Cold denaturation induces inversion of dipole and spin transfer in chiral peptide monolayers. *Nat. Commun.* **2016**, *7* (1), 10744.
18 Ben Dor, O.; Yochelis, S.; Radko, A.; Vankayala, K.; Capua, E.; Capua, A.; Yang, S.-H.; Baczewski, L. T.; Parkin, S. S. P.; Naaman, R.; Paltiel, Y. Magnetization switching in ferromagnets by adsorbed chiral molecules without current or external magnetic field. *Nat. Commun.* **2017**, *8*, 14567.
19 Al-Bustami, H.; Koplovitz, G.; Primc, D.; Yochelis, S.; Capua, E.; Porath, D.; Naaman, R.; Paltiel, Y. Single Nanoparticle Magnetic Spin Memristor. *Small* **2018**, *14* (30), 1801249.
20 Aragonès, A. C.; Medina, E.; Ferrer-Huerta, M.; Gimeno, N.; Teixidó, M.; Palma, J. L.; Tao, N.; Ugalde, J. M.; Giralt, E.; Díez-Pérez, I.; *et al.* Measuring the Spin-Polarization Power of a Single Chiral Molecule. *Small* **2016**, *13*, 1602519–6.
21 Naaman, R.; Sanchez, L. Low Energy Electron Transmission through Thin-Film Molecular and Biomolecular Solids. *Chem. Rev.* **2007**, *107*, 1553−1579.
22 Carmeli, I.; Skakalova, V.; Naaman, R.; Vager, Z. Magnetization of Chiral Monolayers of Polypeptide: A Possible Source of Magnetism in Some Biological Membranes. *Angew. Chem., Int. Ed.* **2002**, *41*, 761−764.
23 Ray, S. G.; Daube, S. S.; Leitus, G.; Vager, Z.; Naaman, R. Chirality-Induced Spin-Selective Properties of Self-Assembled Mono- layers of DNA on Gold. *Phys. Rev. Lett.* **2006**, *96*, 036101.
24 Mishra, D.; Markus, T. Z.; Naaman, R.; Kettner, M.; Göhler, B.; Zacharias, H.; Friedman, N.; Sheves, M.; Fontanesi, C. Spin-Dependent Electron Transmission Through Bacteriorhodopsin Embedded In Purple Membrane. *Proc. Natl. Acad. Sci.* **2013**, *110*, 14872.
25 Roy, P.; Kantor-Uriel, N.; Mishra, D.; Dutta, S.; Friedman, N.; Sheves, M.; Naaman, R. Spin-Controlled Photoluminescence In Hybrid Nanoparticles Purple Membrane System. *ACS Nano.* **2016**, *10*, 4525.
26 Kiran, V.; Mathew, S. P.; Cohen, S. R.; Hernández Delgado, I.; Lacour, J.; Naaman, R. Helicenes—A New Class Of Organic Spin Filter. *Adv. Mater.* **2016**, *28*, 1957.
27 Aragonès, A. C.; Aravena, D.; Cerdá, J. I.; Acís Castillo, Z.; Li, H.; Real, J. A.; Sanz, F.; Hihath, J.; Ruiz, E.; Díez-Pérez, I. Large Conductance Switching in a Single-Molecule Device through Room Temperature Spin-Dependent Transport. *Nano Lett.* **2016**, *16* (1), 218–226.
28 Aragonès, A. C.; Aravena, D.; Valverde-Muñoz, F. J.; Real, J. A.; Sanz, F.; Díez-Pérez, I.; Ruiz, E. Metal-Controlled Magnetoresistance at Room Temperature in Single-Molecule Devices. *J. Am. Chem. Soc.* **2017**, *139* (16), 5768–5778.
29 Silvaggi, N. R.; Martin, L. J.; Schwalbe, H.; Imperiali, B.; Allen, K. N. Double-Lanthanide-Binding Tags for Macromolecular Crystallographic Structure Determination. *J. Am. Chem. Soc.* **2007**, *129*, 7114–7120.
30 Daughtry, K. D.; Martin, L. J.; Sarraju, A.; Imperiali, B.; Allen, K. N. Tailoring Encodable Lanthanide-Binding Tags as MRI Contrast Agents. *Chembiochem Eur. J. Chem. Biol.* **2012**, *13*, 2567–2574.
31 Nitz, M.; Franz, K. J.; Maglathlin, R. L.; Imperiali, B. A Powerful Combinatorial Screen to Identify High-Affinity terbium(III)-Binding Peptides. *Chembiochem Eur. J. Chem. Biol.* **2003**, *4*, 272–276.
32 Nitz, M.; Sherawat, M.; Franz, K. J.; Peisach, E.; Allen, K. N.; Imperiali, B. Structural Origin of the High Affinity of a Chemically Evolved Lanthanide-Binding Peptide. *Angew. Chem. Int. Ed.* **2004**, *43*, 3682–3685.
33 Franz, K. J.; Nitz, M.; Imperiali, B. Lanthanide-Binding Tags as Versatile Protein Coexpression Probes. *Chembiochem Eur. J. Chem. Biol.* **2003**, *4*, 265–271.
34 Sek, S.; Misicka, A.; Swiatek, K.; Maicka, E. J. Conductance of α-Helical Peptides Trapped within Molecular Junctions *Phys. Chem. B* **2006**, *110*, 19671.
35 Rosaleny, L. E.; Cardona-Serra, S.; Escalera-Moreno, L.; Baldoví, J. J.; Gołębiewska, V.; Wlazło, K.; Casino, P.; Prima-García, H.; Gaita-Ariño, A.; Coronado, E. Peptides as Versatile Platforms for Quantum Computing. *J. Phys. Chem. Lett.* **2018**, *9*, 4522–4526.
36 Rosaleny L.E.; Zinovjev, K.; Tuñón, I.; Gaita-Ariño A. A first peek into sub-picosecond dynamics of spin energy levels in magnetic biomolecules Phys. Chem. Chem. Phys., **2019**, *21*, 10908-10913
37 Mondal, P. C.; Fontanesi, C.; Waldeck, D. H.; Naaman, R. Spin-Dependent Transport through Chiral Molecules Studied by Spin-Dependent Electrochemistry. *Acc. Chem. Res.* **2016**, *49*, 2560–2568.
38 Barb, A. W.; Ho, T. G.; Flanagan-Steet, H.; Prestegard, J. H. Lanthanide binding and IgG affinity construct: Potential applications in solution NMR, MRI, and luminescence microscopy. *Protein Sci.* **2012**, *21* (10), 1456–1466.
39 Bard, A. J.; Faulkner, L. R. *Electrochemical Methods: Fundamentals and Application, 2nd Edition*; Wiley, New York, **2001**.
40 Naaman, R.; Waldeck, D. H. Chiral-induced spin selectivity effect. *J. Phys. Chem. Lett.* **2012**, *3* (16), 2178–2187.

41 Simeone, F. C.; Yoon, J. Y.; Thuo, M. M.; Barber, J. R.; Smith, B.; Whitesides, G. M. Defining the Value of Injection Current and Effective Electrical Contact Area for EGaIn-Based Molecular Tunneling Junctions. *J. Am. Chem. Soc.* **2013**, *48*, 18131-18144.

42 Bâldea, I. Transition Voltage Spectroscopy: An Appealing Tool of Investigation in Molecular Electronics, in: Molecular Electronics: an Experimental and Theoretical Approach, **2016**: pp. 397–437

43 Mondal, P. C.; Roy, P.; Kim, D.; Fullerton, E. E.; Cohen, H.; Naaman, R. Photospintronics: Magnetic Field-Controlled Photoemission and Light-Controlled Spin Transport in Hybrid Chiral Oligopeptide-Nanoparticle Structures *Nano Lett.* **2016**, *16* (4), 2806–2811.

44 Kettner, M.; Göhler, B.; Zacharias, H.; Mishra, D.; Kiran, V.; Naaman, R.; Fontanesi, C.; Waldeck, D. H.; Sek, S.; Pawowski, J.; Juhaniewicz, J. Spin Filtering in Electron Transport Through Chiral Oligopeptides. *J. Phys. Chem. C* **2015**, *119* (26), 14542–14547.

45 Soler, J. M.; Artacho, E.; Gale, J. D.; Garcia, A.; Junquera, J.; Ordejon, P.; Sanchez-Portal, D. The SIESTA method for ab initio order-N materials. *J. Phys.: Cond. Matt.* **2002**, *14*, 2745.

46 Rocha, A. R.; García-Suárez, V. M.; Bailey, S.; Lambert, C.; Ferrer, J.; Sanvito, S. Spin and molecular electronics in atomically generated orbital landscapes. *Phys. Rev. B*, **2006**, *73* (8), 085414.

47 Poneti, G.; Bernot, K.; Bogani, L.; Caneschi, A.; Sessoli, R.; Wernsdorfer, W.; Gatteschi, D. A rational approach to the modulation of the dynamics of the magnetisation in a dysprosium–nitronyl-nitroxide radical complex. *Chem. Commun.* **2007**, 1807–1809.

48 Benelli, C.; Caneschi, A.; Gatteschi, D.; Pardi, L. Gadolinium (III) complexes with pyridine-substituted nitronyl nitroxide radicals. *Inorg. Chem.* **1992**, *31*, 741–746.

49 Xu, J. X.; Ma, Y.; Xu, G. F.; Wang, C.; Liao, D. Z.; Jiang, Z. H.; Yan, S. P.; Li, L. C. A four-spin ring with alternating magnetic interactions formed by pyridine-substituted nitronyl nitroxide radicals and Gd(III) ions: Crystal structure and magnetic properties. *Inorg. Chem. Commun.* **2008**, *11*, 1356–1358.

50 Reis, S. G.; Briganti, M.; Soriano, S.; Guedes, G. P.; Calancea, S.; Tiseanu, C.; Novak, M. A.; del Águila-Sánchez, M. A.; Totti, F.; Lopez-Ortiz, F.; Andruh, M.; Vaz, M. G. F. Binuclear Lanthanide-Radical Complexes Featuring Two Centers with Different Magnetic and Luminescence Properties. *Inorg. Chem.* **2016**, *55* (22), 11676–11684.

# Reinforced room temperature spin filtering in chiral paramagnetic metallopeptides

Ramón Torres-Cavanillas, Garin Escorcia-Ariza, Isaac Brotons-Alcázar, Roger Sanchís-Gual, Prakash Chandra Mondal, Lorena E. Rosaleny, Silvia Giménez-Santamarina, Michele Sessolo, Marta Galbiati, Sergio Tatay, Alejandro Gaita-Ariño[*], Alicia Forment-Aliaga[*], Salvador Cardona-Serra[*]

ICMol. Universitat de València. C/ Catedrático José Beltrán nº 2, 46980 Paterna, Valencia, España.

**Table of contents**

# S1 Preparation and characterization of the self-assembled monolayers (SAMs)

**Reagents**

Dimethylformamide, $TbCl_3$, $YCl_3$, 1-octadecanethiol and Gallium Indium Eutectic (Sigma-Aldrich) and Ethanol (Honeywell), are commercially available and were used without any further purification. Poly(L-alanine) with sequence [$SHCH_2CH_2CO$-{Ala-Aib}$_8$-COOH] was bought to Genemed Synthesis, Inc. and the lanthanide binding tag of sequence YIDTNNDGWYEGDELC (LBTC) was purchased from GenScript USA Inc.

**-TbLBTC SAM on Au substrates (sample 1)**

As support we used thermally evaporated Au on $Si/SiO_2$ substrate activated by freshly prepared "piranha solution" (*composition: $H_2SO_4$: $H_2O_2$ (7:3, v/v); Caution: piranha solution is an extremely strong oxidizing agent and should be handled with special attention*), then the substrate was copiously rinsed with deionized (DI) water and dried under Ar.

We followed a self-assembling method to graft TbLBTC from solution on freshly cleaned Au substrates. The substrate was incubated overnight in a buffered TbLBTC aqueous solution formed by: deoxygenated HEPES 10 mM pH 7.0, NaCl 100 mM (from now called buffer), $TbCl_3$ 0.11 mM, TCEP 0.5 mM and LBTC 0.1 mM. In order to remove non-chemisorbed material that could remain on the surface, the sample was copiously rinsed with DI water.

Incubation of the substrates in a solution formed by the target molecule and the solvent is a standard approach for the formation of alkanethiol self-assembled monolayers (SAMs) on metallic surfaces. However, due to the peptidic nature of LBTC, it was also necessary to control the media pH and maintain reducing conditions to assure its stability. Thus, aqueous buffered solutions of LBTC were prepared which contained 4-(2-hydroxyethyl)-1-piperazineethanesulfonic acid (HEPES) as a buffer, a reducing agent and, when required, a lanthanide or yttrium chloride. One of the more common reducing agents used to avoid the oxidation of thiol groups in proteins is 2-mercaptoethanol. However, because of the possible competence of thiol groups from this reducing afent with the cysteine thiol group in the LBTC when the peptide was to form the sulfur-gold bond during the assembling process on the metallic surface, a reducing agent with no thiol groups was selected. Thus, tris(2-carboxyethyl) phosphine (TCEP) was used as reducing agent during the preparation of the LBTC solutions. This phosphine can keep the cysteine from forming disulfide bonds (oxidation problems), but it has not a special affinity for the Au surface. On the other hand, the use of TCEP can mean a different problem. When a metallic cation like $Tb^{3+}$ is added to the solution to form TbLBTC complex, the possibility of a chelating effect of the TCEP (which contains three carboxylic groups), cannot be ruled out. To guarantee that LBTC coordinates the $Ln^{3+}$ cation qualitatively in presence of TCEP, we always worked with an excess of the lanthanide and we checked the luminescence spectra of solutions containing LBTC, TCEP and $Ln^{3+}$ show distinctive signals arising from the successful formation of LnLBTC under these conditions (Fig. S1.1). Therefore the presence of luminescence demonstrates the coordination, as the luminiscence of TbLBTC compounds originates from the sensitization of $Tb^{3+}$ due to the strategically located tryptophan (indole) on the sequence when $Tb^{3+}$ is coordinated by the peptide.[1]

---

[1] Allen, K. N.; Imperiali, B. *Curr. Opin. Chem. Biol.* **2010**, *14* (2), 247–254.

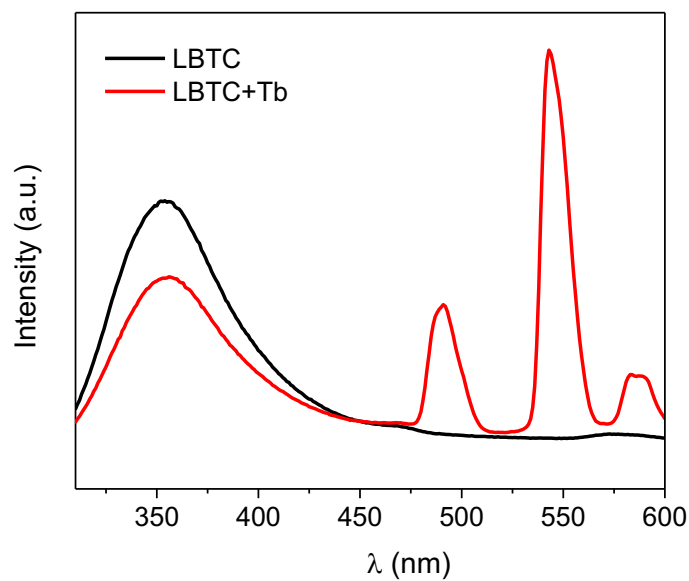

**Figure S1.1.** Luminescence spectra (emission spectra upon 280 nm excitation) of 0.1 mM LBTC (black) and 0.1 mM TbLBTC (0.1 mM LBTC, 0.11 mM $TbCl_3$)(red) in a 0.5 mM TCEP, 10 mM HEPES pH 7.0 and 100 mM NaCl buffer.

-**TbLBTC SAM on Ni/Au substrates (sample 2)**

A similar methodology than the one used with Au substrates was carried out for substrates with 200 nm of ferromagnetic Ni covered with 5 nm of Au (both thermally evaporated on $Si/SiO_2$ substrates). The only difference was that the piranha treatment was substituted by $H_2$ plasma treatment (at~0.15 mbar, 30 sccm for 2 min) to avoid the oxidation of the Ni.

-**YLBTC SAM on a Ni/Au substrate (sample 3)**

An analogous protocol to the one developed for the TbLBTC SAM on Ni/Au was carried out for the Yttrium compound. The Ni/Au substrate was cleaned by a $H_2$ plasma treatment, followed by its immersion in a TCEP 0.5mM, HEPES 10 mM, NaCl 100 mM, Tb 0.11 mM, LBTC 0.1 mM deoxygenated water solution overnight. Finally, to remove non-chemisorbed material that could remain on the surface, the sample was copiously rinsed with DI water.

-**Octadecanethiol (C18S) SAM on a Ni/Au substrate (sample 4)**

The magnetic substrate was immersed in a 10 mM solution of C18S in ethanol overnight. Next, the substrate was copiously rinsed with ethanol. Water contact angle (WCA) measurements were carried out to evaluate the hydrophobicity of functionalized surfaces. Thus, WCA values higher than 110° prove the quality of the resulting SAM (Fig. S5).

- **Poly(L-alanine) with 8 aminoacid units (Ala8) SAM on a Ni/Au substrate (sample 5)**

A freshly prepared Ni/Au substrate was cleaned in boiling acetone during 15 min, followed by a 15 min washing step in ethanol. Next, the substrate was plasma treated with $H_2$ (at~0.15 mbar, 30 sccm for 2 min) and incubated for 48 h in an Ala8 0.1mM DMF solution. Finally, the SAM modified substrate was rinsed with clean DMF and dried under a $N_2$ stream.

**Photoluminescence measurements**

To prove that $Tb^{3+}$ was still coordinated by the LBTC unit when molecules are attached to the metallic surface, reflectance photoluminescence measurements were carried out in an optical microscope NIKON Eclipse LV-100, equipped with a digital camera: Nikon, D7000 AF-S DX NiKKOR 18-105mm f/3.5-5.6G ED VR. An UV lamp was used as a source of the 254 nm excitation wavelength. Photoluminescence intensity was recorded by the digital camera of the optical microscope and quantified from the integrated density of pixels through the ImageJ software.

Solutions of LBTC, $Tb^{3+}$, YLBTC, and TbLBTC were measured to validate the homemade setup. As can be observed in figure S1.2a, only TbLBTC solution originates the expected photoluminescence, which is recorded by the digital camera of the optical microscope. A baseline is observed for the other samples.

In the same way, four different substrates were evaluated for comparison: Clean Au (Au), Au incubated in $Tb^{3+}$ solution and copiously rinsed (Au/Tb), Au with a SAM of YLBTC (Au/YLBTC) and Au with a SAM of TbLBTC (Au/TbLBTC). It should be noted for these surface photoluminescence measurements, the reduced amount of material and the increased background signal coming from the light reflection on the substrate, leads to smaller differences between the baseline and the signal. However, as in the previous case only Au/YLBTC displays luminescence compared with the control samples. Supporting our claim that $Tb^{3+}$ is indeed coordinated by the LBTC peptide also on the surface, and consequently the peptide is likely to maintain the folding that allows for the lanthanide binding.

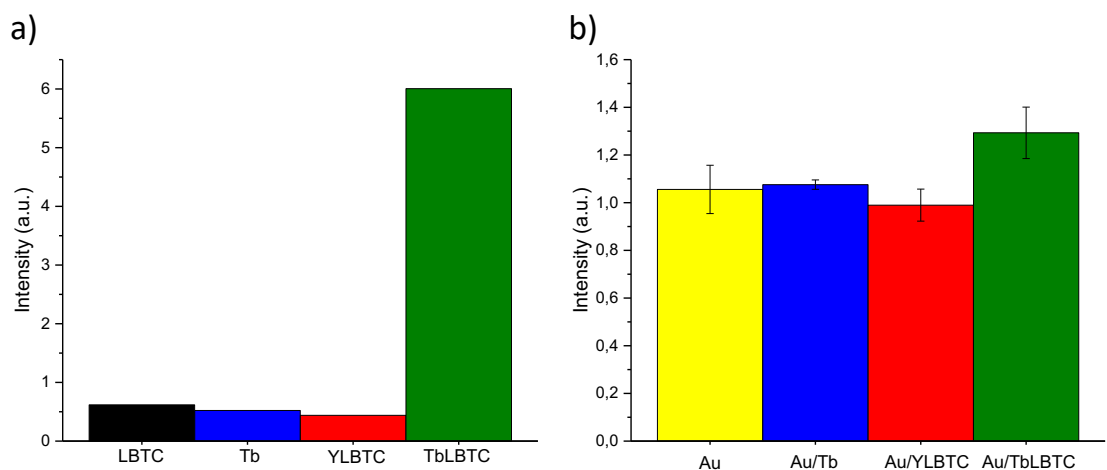

**Figure S1.2** Photoluminscense signal recorded by the digital camera of the optical microscope (a) for solutions of LBTC, $Tb^{3+}$, YLBTC and TbLBTC and (b) for clean Au (Au), Au incubated in $Tb^{3+}$ solution and copiously rinsed (Au/Tb), Au with a SAM of YLBTC (Au/YLBTC) and Au with a SAM of TbLBTC (Au/TbLBTC).

# S2 Atomic force microscopy (AFM)

The topography of the different functionalized substrates was imaged with a Digital Instruments Veeco Nanoscope IVa Atomic Force Microscope in tapping and contact mode, using silicon tips with natural resonance frequency of 300 kHz and with an equivalent constant force of ~40 N/m. AFM images of the substrates after functionalization show the absence of huge aggregates and that substrate roughness is not increased. (Fig. S2.1 and S2.2)

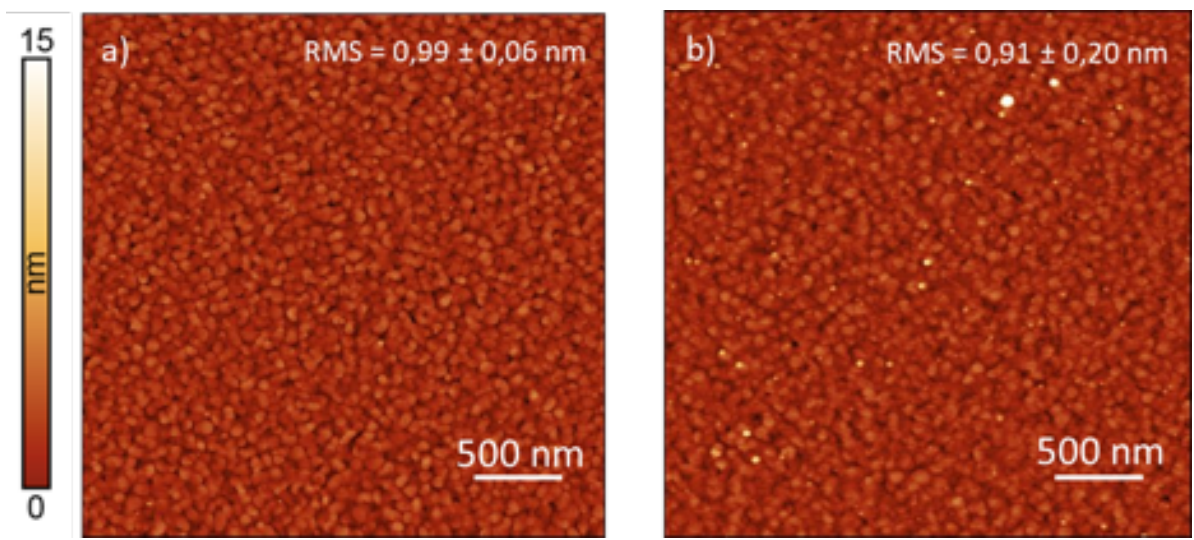

**Figure S2.1** (a,b) Atomic force microscopy images of a Au surface incubated overnight with buffered solution, including $Tb^{3+}$, in the absence (a), or presence of 0.1 mM LBTC (b).

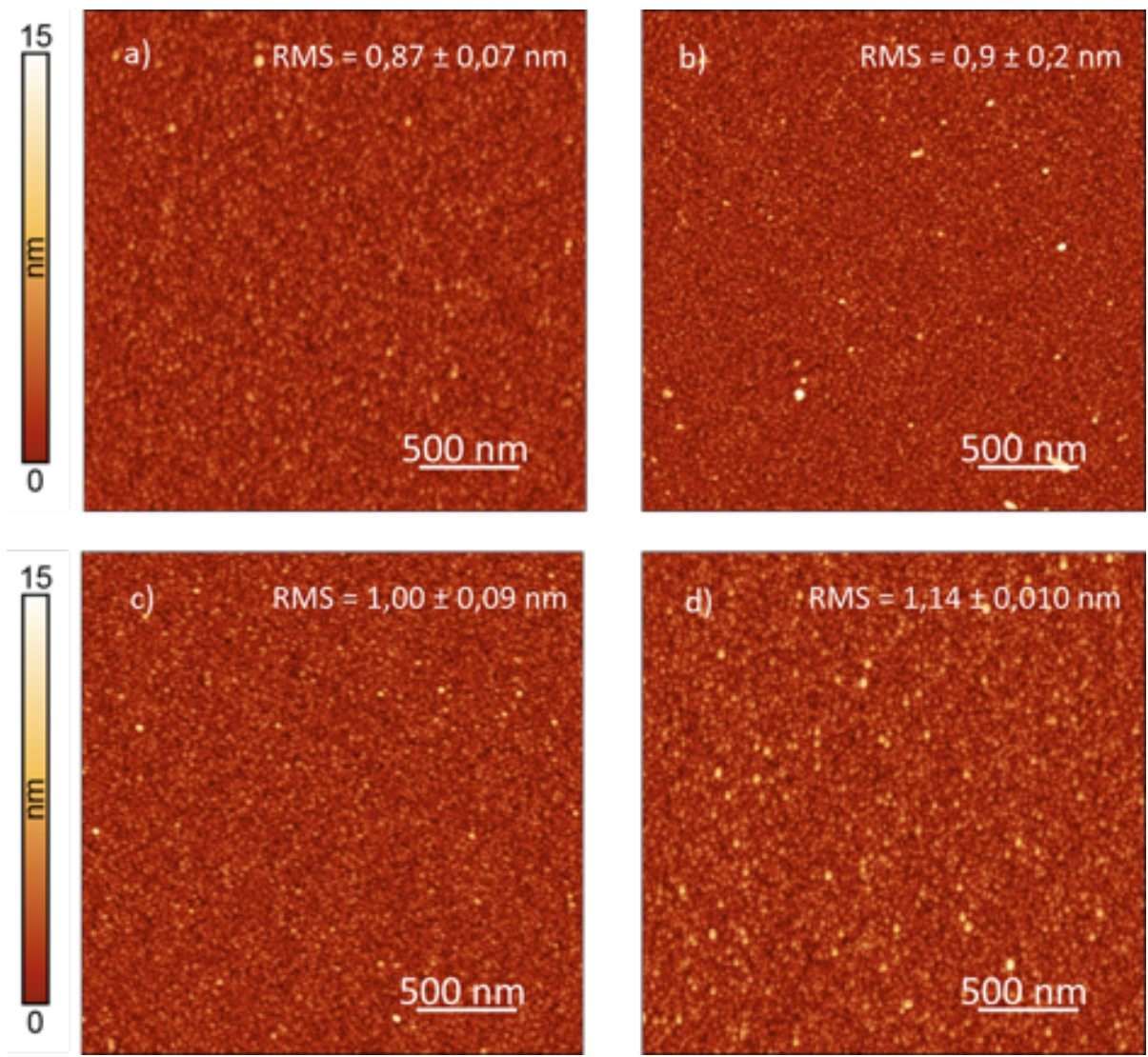

**Figure S2.2** (a-d) Atomic force microscopy images of a Ni/Au surfaces incubated overnight in buffer solution (a), with Ala8 (b) and with 0.1 mM LBTC coordinating $Tb^{3+}$ and $Y^{3+}$ (c and d, respectively).

# S3 Matrix-assisted laser desorption/ionization-time of flight (MALDI-TOF)

TbLBTC/Au samples were analyzed in a 5800 MALDI TOF instrument (ABSciex) in positive reflectron mode (3000 shots every position) in a mass range of 150–3000 m/z. Previously, the plate and the acquisition method were calibrated with a CalMix solution. The analysis was carried out at the Proteomics Unit in the SCIE of the Universitat de València. MALDI is a soft ionization technique commonly used for the mass spectrometry analysis of bio- and large organic molecules. The spectrum shows charged fragments of the macromolecule that are obtained in the gas phase after ionizing the sample. Soft interactions like electrostatic ones between LBTC and Ln are easily broken during this ionization process, and only LBTC fragments (charge +1) are expected in the mass spectra. As shown in Fig. S3, the main fragment of the SAM appears at 1630 m/z; there is also a smaller peak at 1929 m/z. The peak at 1630 m/z belongs to the fragment DTNNDGWYEGDELC and the peak at 1929 m/z corresponds to the molecular weight of the LBTC with an additional $Na^+$. Thus, the MALDI-TOF result ensured the formation of the SAM, preserving the integrity of the molecule after its attachment on the Au surface.

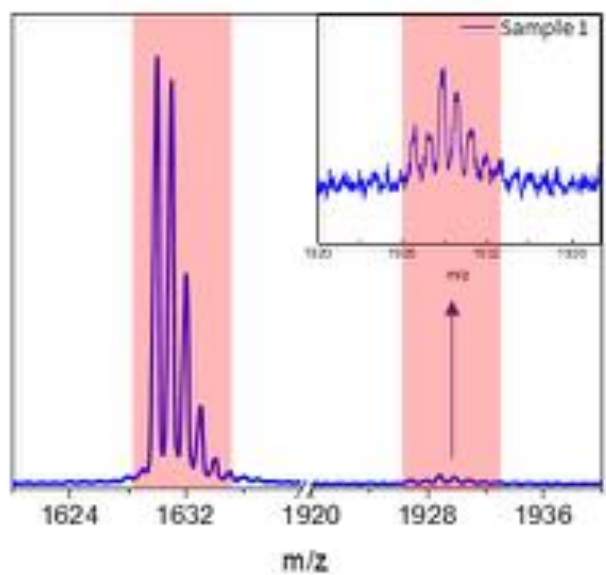

**Figure S3.** MALDI-TOF spectrum of TbLBTC SAM on a Au surface.

# S4 X-ray photoemission spectroscopy (XPS)

X-ray photoelectron spectroscopy (K-ALPHA, Thermo Scientific) was used to analyze the surfaces of the samples. Samples were analyzed ex-situ at the X-ray Spectroscopy Service at the Universidad de Alicante. All spectra were collected using Al Kα radiation (1486.6 eV), monochromatized by a twin crystal monochromator, yielding a focused X-ray spot (elliptical in shape with a major axis length of 400 μm) at 3 mA·C and 12 kV. The alpha hemispherical analyzer was operated in the constant energy mode with survey scan pass energies of 200 eV to measure the whole energy band and 50 eV in a narrow scan to selectively measure the particular elements. Spectra are referenced using the C 1s main peak (284.8 eV).

XPS spectra of TbLBTC SAMs show the presence of Tb3d peaks (1277 and 1242 eV), denoting the signature of the lanthanide on the surface. Due to the extremely high affinity of the LBT series of compounds for $Tb^{3+}$ ions ($K_D$ ~ 5-10 nM)[2] it is fair enough to assume that the observed Tb ions are bounded to the LBTC. This particular point has already been addressed in section S1). The presence of the binding tag is also probed by S2p (162 eV, thiolate bonded to Au) and N1s (400 eV, amino groups) (Fig. S4.1).

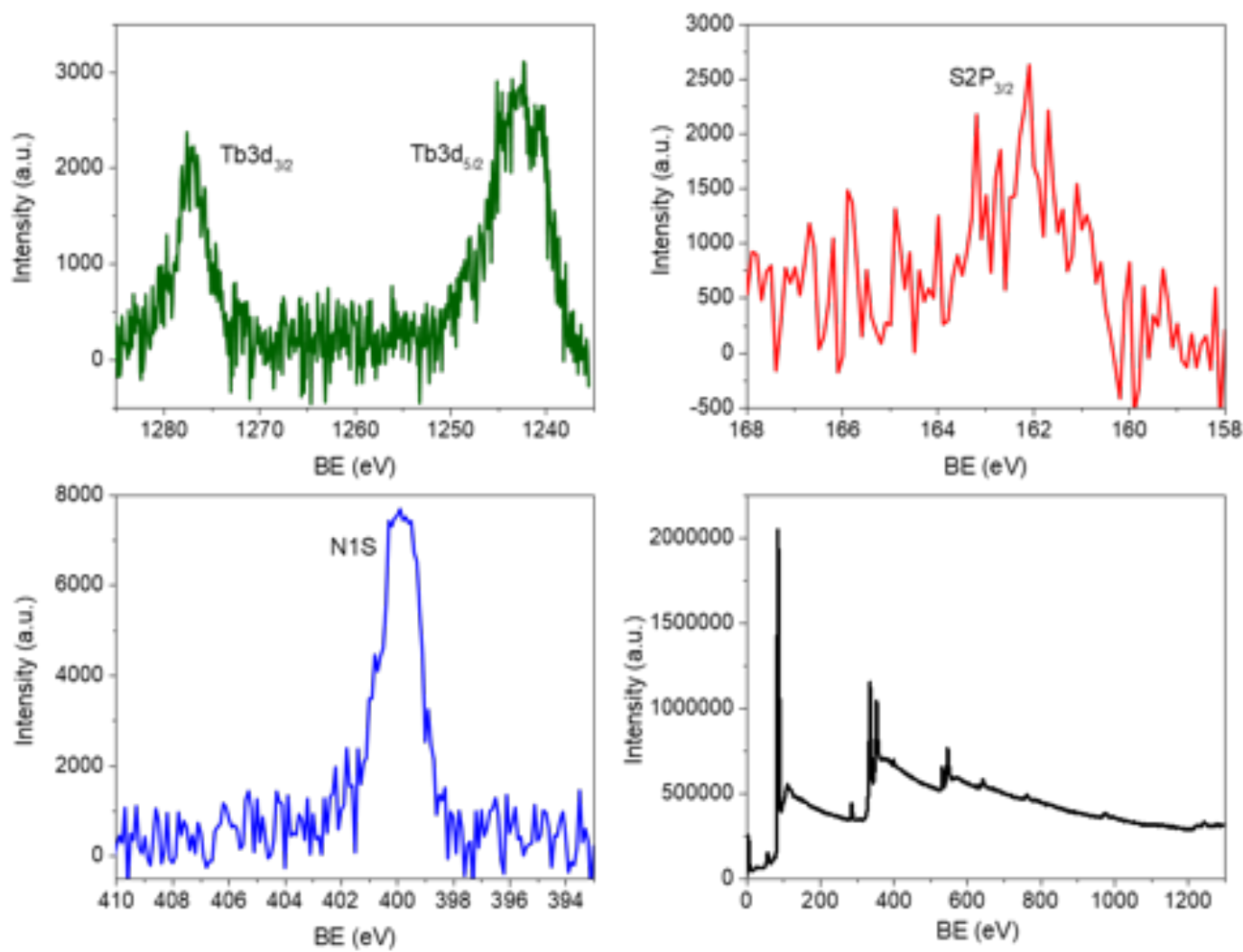

**Figure S4.1.** XPS spectrum of Tb (green), S (red) and N (blue) and survey (black) of TbLBTC SAM on a Au surface.

---

[2] Martin, L. J.; Sculimbrene, B. R.; Nitz, M.; Imperiali, B. *QSAR Comb. Sci.* **2005**, *24* (10), 1149–1157.

## S5 C18S contact angle measurements

Dynamic water contact angle measurements of the samples were performed in air using a Ramé-hart 200 standard goniometer equipped with an automated dispensing system. The initial drop volume was 0.17 µL, increased by additions of 0.08 µL and waiting times of 1500 ms for each step. The high values displayed for sample 4 (>110º) indicate the good quality of the resulting C18S SAM (See Fig. S5).

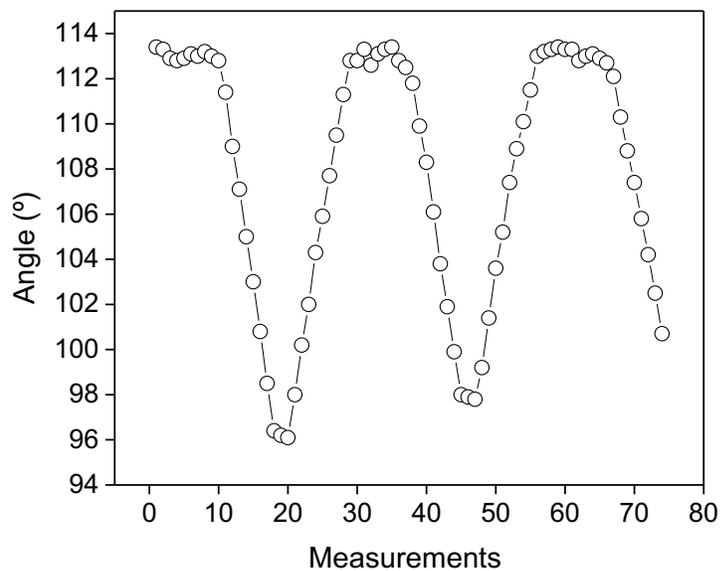

**Figure S5.** Contact angle measurement of sample 4**.**

# S6 Coverage evaluation using quartz crystal microbalance (QCM)

In this work, for a typical experiment, a freshly cleaned Au covered piezoelectric quartz crystal was immersed in a liquid cell where a flow of different liquids at controlled rates were pumped while the oscillation frequency of the crystal was continuously monitored. Several cycles of pure water and aqueous buffered solution were performed until the frequency was stabilized (Fig. S6.1, water in red, buffer in green). Right after a buffered solution step, TbLBTC or LBTC solution was injected during ca. 60 min to ensure a constant frequency value. Finally, a flow of the buffered aqueous solution was introduced to remove physisorbed TbLBTC or LBTC on the crystal.

Sauerbrey's equation was used to estimate the relation between resonant frequency and mass loading of electrodes[3].

$$\Delta f = -\frac{2f_0^2}{A\sqrt{\rho\mu}}\Delta m$$

where $f_0$ is the resonant frequency of QCM, $\Delta f$ is the change in the resonant frequency of quartz microbalance due to the mass change ($\Delta m$) of electrodes, $A$ is the active area, $\rho$ is the density and $\mu$ is the shear modulus of the quartz crystal provided by the manufacturer.

To calculate the theoretical coverage of the surface, two different models have been used that give rise to a range of coverage values (Fig. S6.2):

-Model A: the molecule is considered as a cube of 1.7 nm side in a square arrangement (maximum coverage = 1).
-Model B: the molecule is considered a sphere which projection on the surface is a circle of 1.6 nm diameter, with a compact hexagonal arrangement, giving rise to a maximum coverage density of $\pi/2\sqrt{3}$ (ca. 0.9069).

It is important to clarify that the molecular weights of the LBTC with (2065 g·mol$^{-1}$) or without Tb$^{3+}$ (1906 g·mol$^{-1}$), are not precise as the number and kind of counter ions and amount of coordinating water molecules are unknown. We have considered the simplest situation where protons compensate for negative charges of the polypeptide; this means that the final coverages estimated are the highest values that can be expected.

The coverage calculated from TbLBTC experiments range between 80-116%. On the other side, in spite of the absence of lanthanide in the LBTC, its tertiary structure keeps its round shape (based on LBTC crystal structure), and the same theoretical shape can be estimated. As observed in Fig. S6.1, when LBTC is anchored on the electrode, the frequency shift measured is one third higher than for TbLBTC (ca. 30Hz vs. 20Hz). This large value indicates coverage between 145-189%.

From the analysis of these results, we believe that when Tb complex is formed in solution, the globular-like structure remains unchanged while the thiol groups are attached to the gold and almost all the surface is covered. However, the large amount of material estimated when the Tb$^{3+}$ is not present in the solution, only can be justified by two different explanations: (i) intermolecular polypeptide-polypeptide interactions are stronger than intramolecular interactions forcing a less globular shape and permitting a more compact packing at the surface, or (ii) a second layer is formed on top of the first layer. From our point of view, and in accordance with the film thickness determined in section S7, the second option is less probable as there is no reason to think that a second layer can be formed with LBTC, but it does not happen with TbLBTC, so it should be observed in both cases.

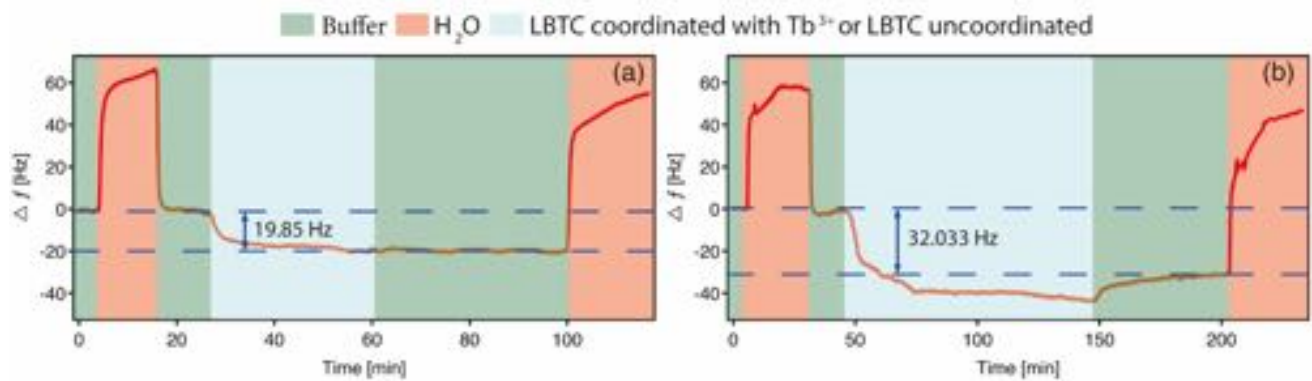

**Figure S6.1.** Quartz crystal microbalance measurements of LBTC coordinating Tb$^{3+}$ (a) and LBTC uncoordinated (b).

---

[3] Sauerbrey, G.; Verwendung von Schwingquarzen zur Wägung dünner Schichten und zur Mikrowägung. *Zeitschrift für Phys*. **1959**, *155*, 206.

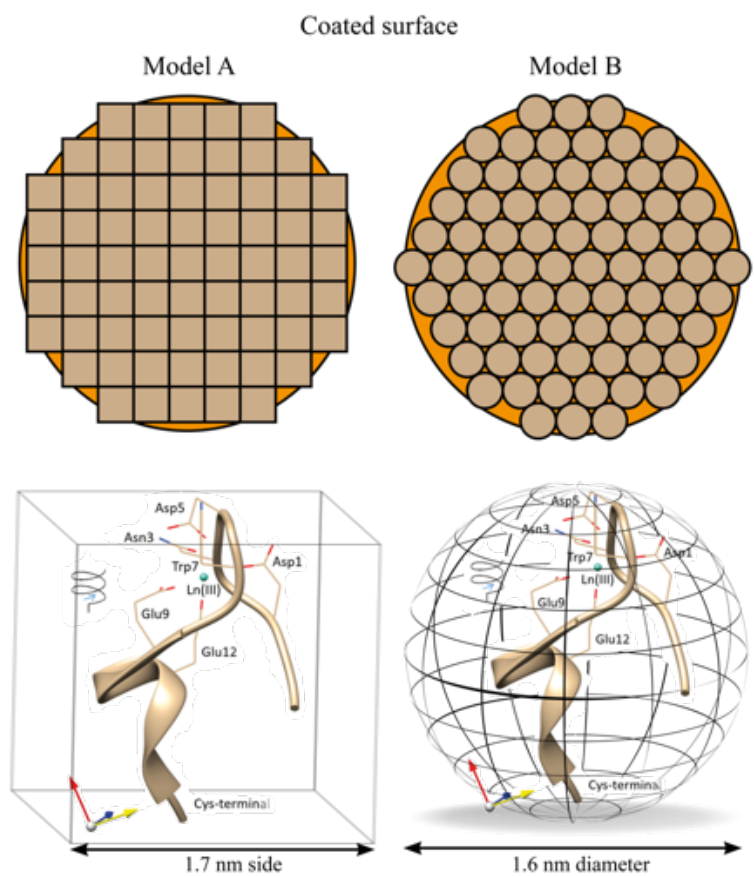

**Figure S6.2.** Area of the base considered to calculate theoretical coverage of the polypeptide according to the two models proposed, Model A on the left and Model B on the right both using crystallographic structure of LBT.

# S7 Film thickness estimation.

Film thickness estimation was accomplished using Spectroscopic Ellipsometry and XPS. Both results agree with the formation of a layer of ca. 1.7 nm. This result agrees with the formation of a single TbLBTC monolayer, where the polypeptide keeps its globular shape (See previous Figure S6.2).

## S7.1 Ellipsometry

Ellipsometry measurements were carried on a GES5E Variable Angle Spectroscopic Ellipsometer with an incident angle of 68º and a spot size of 65 μm x 80 μm in the 1.5-5 eV range. Data was modelled using Winelli 3 software and a two layers model. Clean gold substrates (200 nm Au thermally evaporated on silicon) were modelled first and used as a reference. Next, a Cauchy model (A = 1.5) was used to model the organic film in the Au/TbLBTC sample.[4] Film thickness was found to be approximately 1.85 nm. Best fittings are displayed in Figure S7.1.

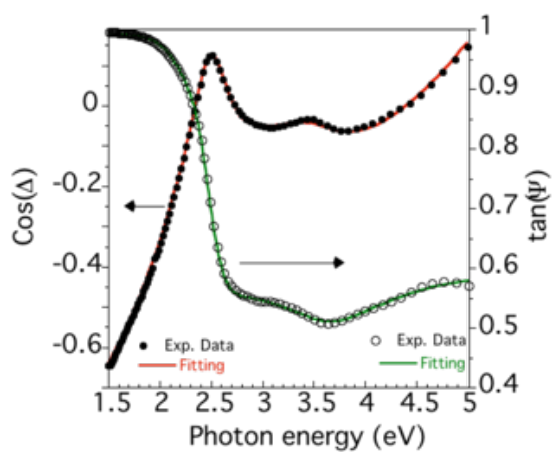

**Figure S7.1** Fits to the ellipsometry data for a TbLBTC SAM on a gold substrate.

## S7.2 XPS study of TbLBTC SAM thickness

The simplest approach to determine overlayer thicknesses assumes that an increasing overlayer thickness causes an exponential attenuation of the photoelectron signal coming from the substrate that can be described by:

$$I = I_0 \exp(-d/\lambda) \qquad [1]$$

Where $I_0$ is the signal coming from the substrate in the absence of the overlayer, $d$ is the overlayer thickness and $\lambda$ the inelastic mean free path (IMFP) of the photoelectrons crossing the overlayer.

$I$ and $I_0$ where determined using the same spectrometer model than the one described in section S4 by measuring the intensity of the Au $4f_{7/2}$ peak at 84.4 eV on a TbLBTC/Au sample before and after removing the TbLBTC SAM. To remove the SAM successive Ar ion gun (1000 eV and 30 s) etching cycles were used until and invariable Au signal was obtained.

---

[4] Tompkins, H.G.; McGahan, W.A.; *Spectroscopic Ellipsometry and Reflectometry: A User's Guide*

Equation 1 was then used to calculate the thickness of the TbLBTC SAM (d = 1.7 nm). A value of $\lambda$ = 58.1 Å was estimated from the NIST Electron Inelastic-Mean-Free-Database, using the Greis equation,[5] which utilizes the molecular formula of the compound and its density. The molecular formula used was that of the LBTC peptide coordinating one $Tb^{3+}$ and two $Na^+$ ions, for charge neutralization ($TbNa_2C_{82}H_{111}N_{19}O_{32}$). The density of the monolayer, ~ 1,1 g·cm$^{-3}$, was extracted from de QCM measurements, assuming a full coverage of the substrate.

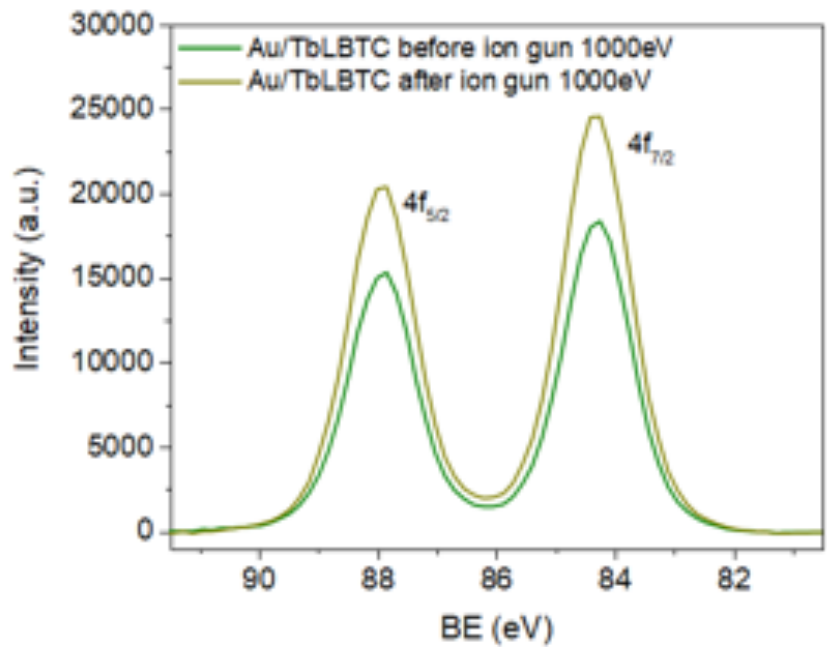

**Figure S7.2.** Au 4f XPS spectra of TbLBTC SAM on a Au surface, before (dark green) and after (light green) ion etching.

---

[5] Gries, W. H. (1996). A Universal Predictive Equation for the Inelastic Mean Free Pathlengths of X-ray Photoelectrons and Auger Electrons. *Surface and Interface Analysis: An International Journal devoted to the development and application of techniques for the analysis of surfaces, interfaces and thin films*, *24*(1), 38-50.

# S8 Spin-dependent electrochemistry set-up and scan rate dependence measurements.

Room temperature spin-dependent electrochemical measurements were executed using an Autolab electrochemical workstation (Model Autolab-128 N potentiostat/galvanostat) which was linked to a personal computer having Nova 2.1 program. A custom-built electrochemical cell was designed such a way that a Nd permanent magnet (0.35T in direct contact) can be placed underneath the working electrode, and it can be flipped easily from "up" to "down" direction or vice-versa without affecting the setup. The cell equipped with three electrodes was used for electrochemical measurements. SAMs modified or unmodified Ni/Au, Ag/AgCl (3M KCl), and cleaned Pt wire were used as working, reference and counter electrode, respectively. As the metallo-peptide is redox inactive, we used an external redox probe to monitor the spin-dependency on the electro-chemical charge-transfer process. Prior to the voltammogram measurements, freshly prepared 5 mM $K_4Fe(CN)_6/K_3Fe(CN)_6$ solution in HEPES, 100 mM NaCl was deoxygenated in Ar for 30 min. The same set up was used for EIS, but using Gamry 1000E and Gamry 5000E potentiostat/galvanostat controlled by Gamry software. An AC voltage of 10 mV in the frequency range of 100–105 Hz at the OCP was applied. EIS data were analyzed and fitted by means of Gamry Echem Analyst v. 7.07 software.

To demonstrate that, as expected, the redox reactions at the working electrode are controlled by a diffusion process, scan rate dependence measurements were performed. Anodic and cathodic currents as a function of the scan rates and square roots of the scan rates for TbLBTC SAM modified electrode and bare electrode are depicted in Fig. S8.1, and Fig. S8.2, respectively showing a better linearity with the square roots of the scan rates[6].

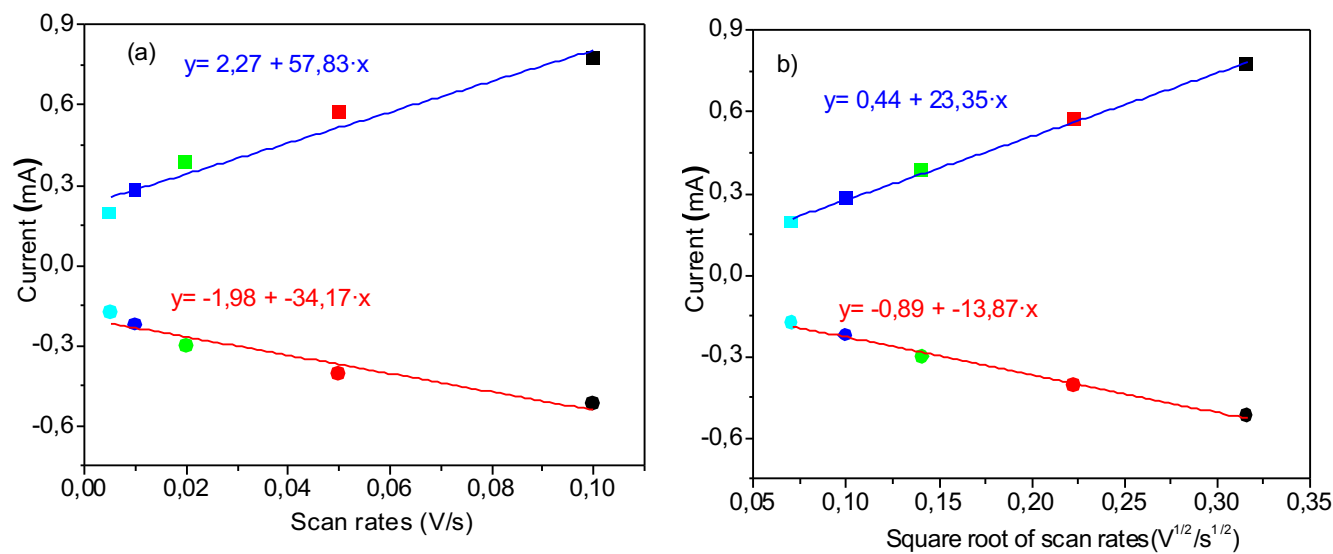

**Figure S8.1.** Plot of Faradaic currents as a function of (a) scan rates (V/s), and (b) square root of scan rates recorded on a TbLBTC modified Ni working electrode. Plot (a) shows less linearity ($R^2$ = 0.91-0.94), while (b) shows more linearity ($R^2$ = 0.996-0.989).

---

[6] Bard, A. J., Faulkner, L. R., Leddy, J., and Zoski, C. G. Electrochemical methods: fundamentals and applications, New York: Wiley, **1980**, *2*.

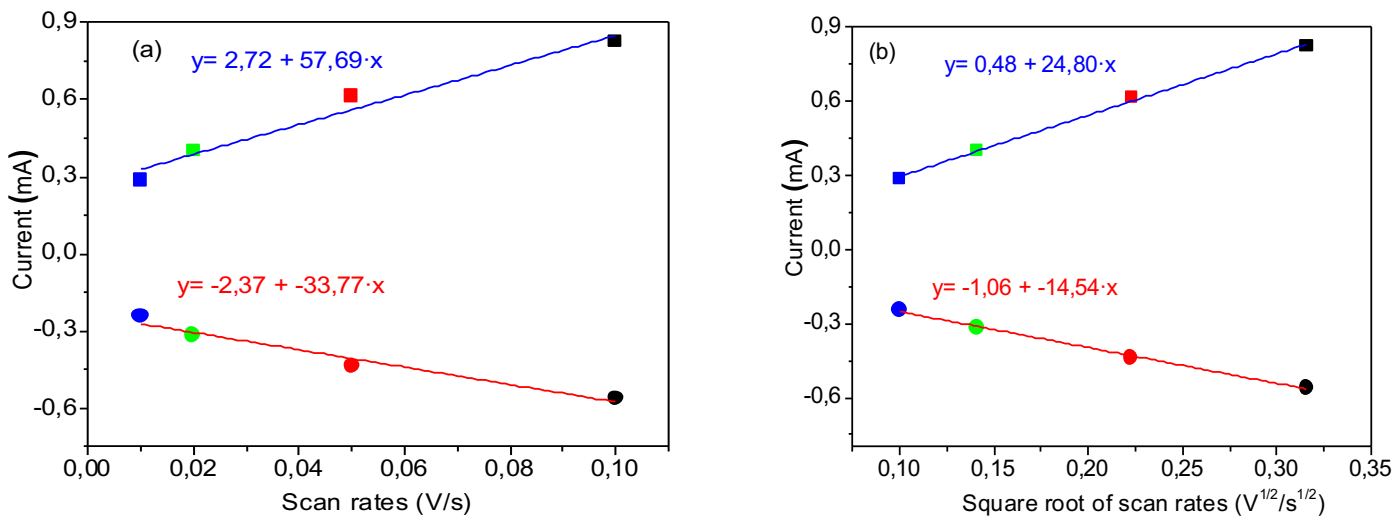

**Figure S8.2.** Plot of Faradaic currents as a function of (a) scan rates (V/s), and (b) square root of scan rates recorded on a bare Ni/Au working electrode. Plot (a) shows less linearity ($R^2$ = 0.95), while (b) shows more linearity ($R^2$ = 0.996-0.997).

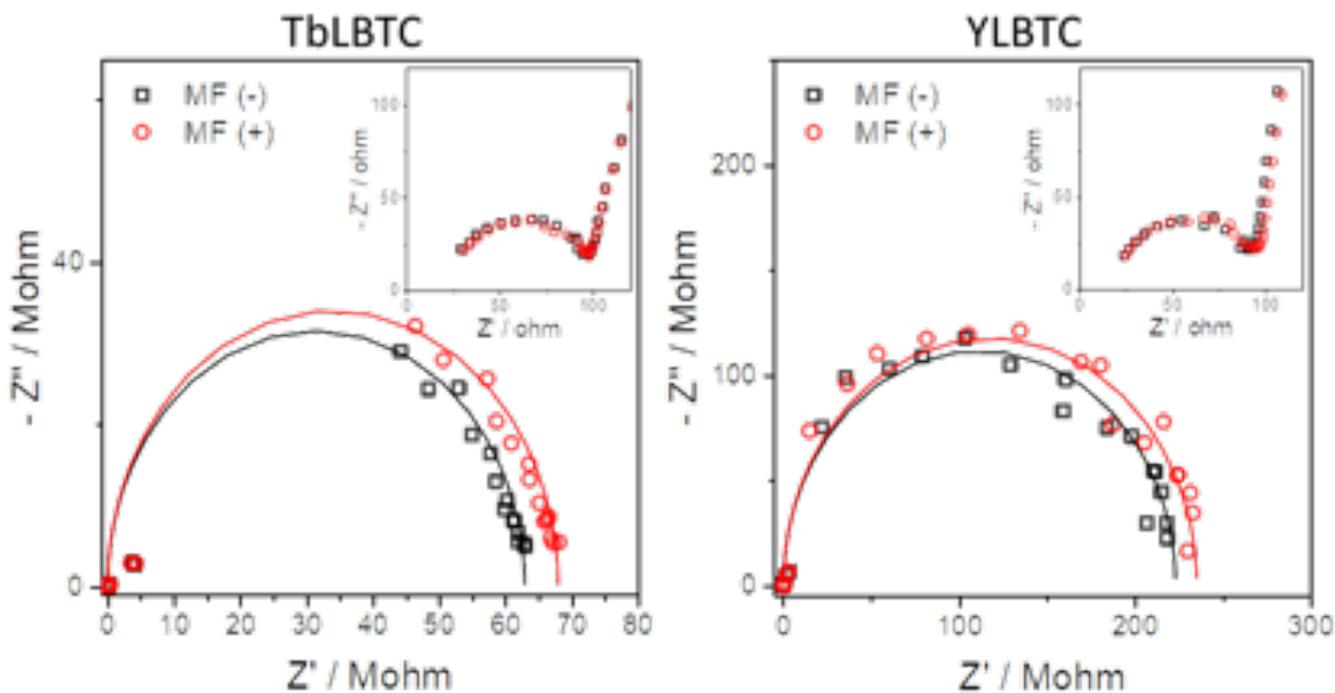

**Figure S8.3.** a) Electrochemical impedance spectroscopy (EIS) of Ni/Au working electrode modified by a SAM of TbLBTC and YLBTC, left and right respectively, under an external magnetic field pointing "up" (red circles) or "down" (black squares).

# S9 EGaIn measurements of SAM films on Au

We measured the current ($I$) as a function of the applied bias ($V$) of the junction formed by bringing a Gallium–Indium eutectic liquid-metal drop into contact with the SAM, using the substrate as the bottom electrode. A Nd permanent magnet (0.35T) was placed in direct contact with the bottom part of the substrate to carry out measurement in the presence of magnetic field. EGaIn drops with diameter of around 0.015 cm was produced using a syringe. The area of the contact ($A$) was estimated from the diameter of the contact between the drop and the SAM using a CCD camera. Data were recorded using a Keithley 6517B electrometer controlled with LabVIEW (National Instruments) with the ground taken from the Eutectic Gallium-Indium drop.

We measured 6 to 22 consecutive $I$-$V$ curves for a given drop size. Next, the sample was displaced laterally with a XYZ-stage, we produced a new Eutectic Gallium-Indium drop, and the whole measurements were repeated all over again. We repeated this process on different parts of the film surface and collected at least 60 I-V curves for each sample.
Current density, $J$ ($J = I/A$) histograms at a given voltage were calculated from the measured $|I(\pm V)|$ values and the area of the contact determined with the CCD camera. The experimental variation in $J$ histogram for a given sample can be attributed to the presence of small local variation on thickness, roughness or structure and/or to errors on the estimation of the drop diameter. As can be seen in Figure 9.1a, compared to the pristine gold samples (whitout SAM) TbBTC/Au samples, the vast majority of the junctions (>90%) showed lower current densities with log ($|J/A \cdot cm^{-2}|$) = ~ -1, roughly three orders of magnitude lower than the pristine sample that displayed log ($|J/A \cdot cm^{-2}|$) = ~ 2, both measured at 0.15V. With less than 10% of contacts given rise to short-circuits: this is the origin of the peak at log($|J(0.15V)|$ ~2. (Fig. 8.1a).

To find the energy level of TbLBTC, we employed the single-level model with Lorentzian transmission (Newns–Anderson model) to analyze the experimental $I$–$V$ curves of sample 1, to estimate the $E_F$ (Fermi energy) -molecular orbital energy offset ($\varepsilon_o = E - E_F$) and the asymmetry of the potential profile ($\gamma$) along the junction. $I$–$V$ curves of the not short-circuited contacts are nearly symmetric and exhibit a slight nonlinearity that becomes more pronounced at higher bias. Fitting of the $I$–$V$ curves up to ±0.5 V yielded $\varepsilon_o$ = 0.68 ± 0.18 eV and $\gamma$ = -0.03 ± 0.03 average values (a representative $I$–$V$ curve and its fitting is shown in Fig. S9.1b). In order to reveal the transition voltage ($V_t$), we collected additional $I$(V) scans up to ±1 V. However, a bias voltage above ±0.5 V resulted in the breakdown of 100% of the junctions after only a few scans. The Fowler−Nordheim plot obtained from a $I-V$ curves measured up to 1 V from the data before junction breakdown is displayed in Fig. S9.1c. It was possible to estimate average $V_{t+}$ and $V_{t-}$ values of 0.85 ± 0.11 V and -0.80 ± 0.08 V (that corresponds to $\varepsilon_o$ = 0.710 eV, $\gamma$ = 0.010) consistent with the previously estimated parameters. The Fowler−Nordheim plot obtained from a $I-V$ curve measured up to 1 V is displayed in Fig. S9.1c. This information combined with the change in the absolute value of the electrode work function ($\Phi$) due to the adsorbed TbLBTC, $\Phi_{SAM}$ = 4.62 eV, led to the energy level alignment scheme of Au/TbLBTC//Ga$_2$O$_3$/EGaIn junction proposed in Fig. S9.2.

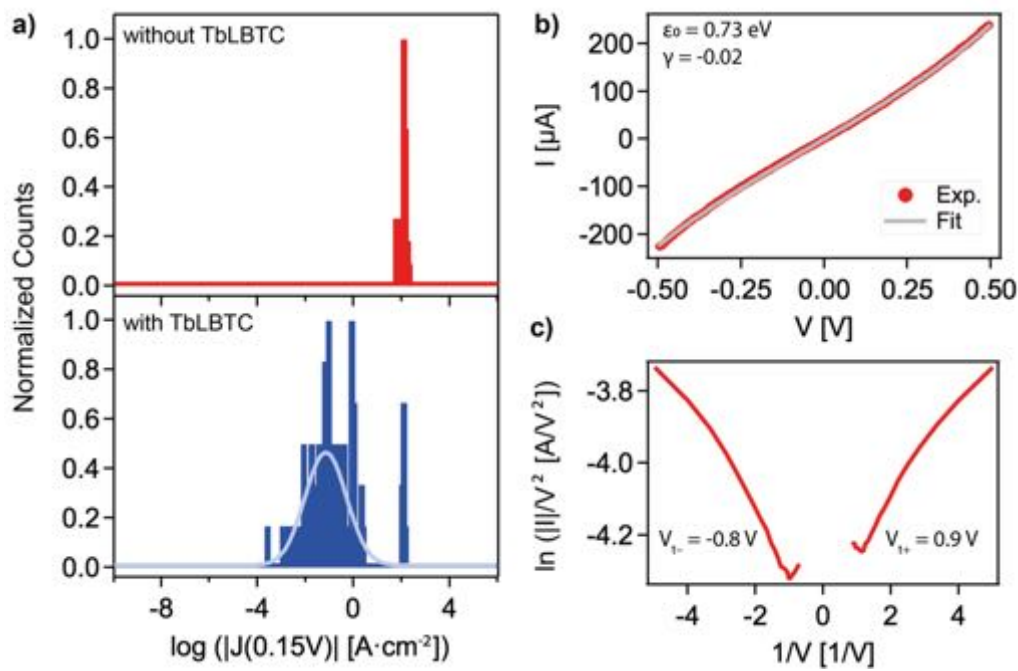

**Figure S9.1.** (a) Histograms of the current density (as log $J$) measured at $V_{bias}$ = 0.15 V on a pristine gold electrode acting as control (top) and on an electrode functionalized with a TbLBTC SAM (bottom). (b) Representative $I–V$ curve for the functionalized electrode, fitted using a single-level model with Lorentzian transmission and (c) Fowler−Nordheim plot obtained from a $I−V$ curve measured up to 1 V.

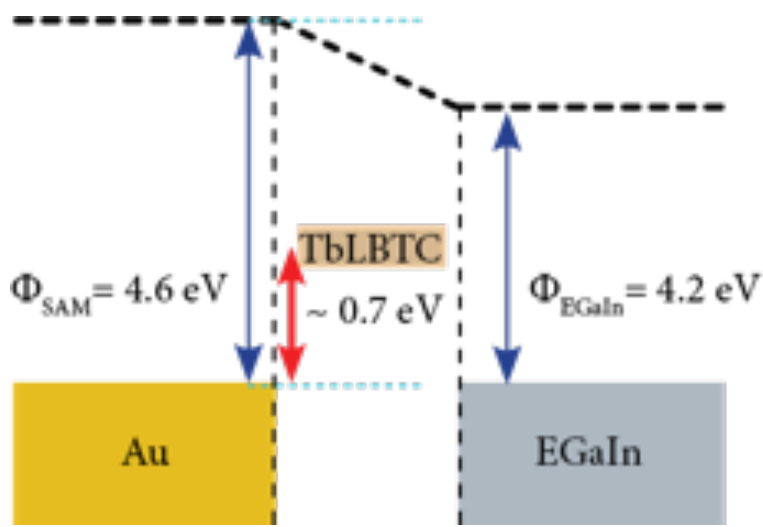

**Figure S9.2.** Schematic energy level alignment diagram of Au/TbLBTC//EGaIn junction. Au work function measured in this work and GaIn eutectic work function obtained from literature.[7]

---

[7] Chiechi, R., Weiss, E., Dickey, M., Whitesides, G. Eutectic Gallium–Indium (EGaIn): A Moldable Liquid Metal for Electrical Characterization of Self-Assembled Monolayers *Angew. Chem. Int. Ed.*, **2008**. *47*, 142 144.

# S10 Kelvin probe and ambient pressure photoemission spectroscopy measurements

Kelvin Probe and Ambient Pressure Photoemission Spectroscopy (APS) were measured using an APS02 system by KP Technology. The system has a Kelvin Probe with a 2 mm metal tip enabling a work function resolution of 1–3 meV. The setup includes a dual-mode APS unit to measure the absolute work function of a material with an excitation range of 3.4 eV to 7.0 eV. The absolute work function of a cleaned bare gold ($\Phi_{Au}$) was estimated by linear extrapolation of a plot of the square root of the photoemission yield as a function of energy (eV), and was found to be 4.96 eV (Fig. S9).

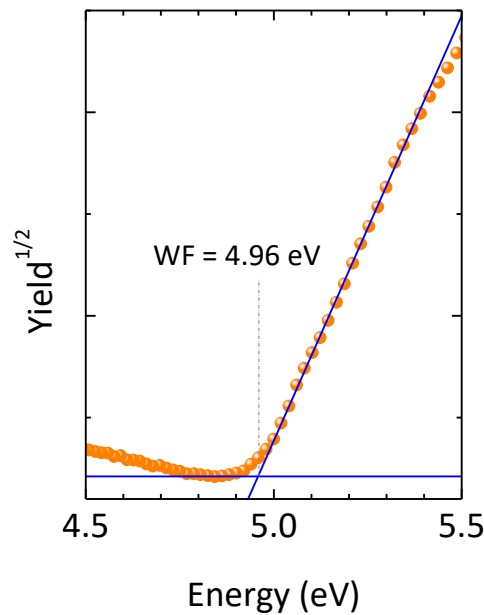

**Figure S10:** APS spectrum of the reference bare gold surface. The work function is estimated by extrapolation of the linear part of the photoemission (square root). E = 0 eV indicates the vacuum level.

Then, the variation in the contact potential difference (CPD) between a bare gold and a TbLBTC-functionalized gold sample (sample 1) was used to estimate the absolute work function of the gold in the presence of the TbLBTC SAM ($\Phi_{SAM}$). The CPD for the gold and SAM-functionalized gold were found to be $0.322 \pm 0.026$ meV and $-0.018 \pm 0.017$ meV, respectively. In order to estimate $\Phi_{SAM} = \Phi_{Tip} + CPD_{SAM}$ (where $\Phi_{Tip}$, is the work function of the KP tip), we measured $\Phi_{Tip} = \Phi_{Au} - CPD_{Au.} = 4.64$ eV. Hence, $\Phi_{SAM}$ was estimated to be 4.62 eV. Similar shifts in work function have been described before for the grafting of thiols on gold.[8]

---

[8] Heimel, G., Romaner, L., Zojer, E., Bredas, J. The Interface Energetics of Self-Assembled Monolayers on Metals. *Accounts of Chemical Research*, **2008**. *41*, 721-729.

# S11 Charge transport calculations on LnLBTC between two gold electrodes

First-principles calculations are performed using the SMEAGOL code that interfaces the non-equilibrium Green's function (NEGF) approach to electron transport with the density functional theory (DFT) package SIESTA.

In our simulations, the transport junction is constructed by placing the polypeptide between two Au (111)-oriented surfaces with a 7x7 cross section. It mimics a standard transport break-junction experiment with the most used gold surface orientation. The choice of a gold electrode arises from the stability of the sulfur-gold bond that ensures the best attachment between the cysteine and the Au surface. The initial S-surface distance was set to 2.0 Å with the S atom located at the 'hollow site', the most stable absorption position discussed in the literature. Thus, the entire structure is then relaxed until the maximum atomic forces are less than 0.01 eV/Å. A real space grid with an equivalent plane wave cutoff of 200 Ry (enough to assure convergence) has been used to calculate the various matrix elements. During the calculation the total system is divided in three parts: a left-hand side leads, a central scattering region (SR) and a right-hand side lead. The scattering region contains the molecule as well as 4 atomic layers of each lead, which are necessary to relax the electrostatic potential to the bulk level of Au. Due to the internal limitations of the code, $Gd^{3+}$ derivative was used instead of $Tb^{3+}$. It is possible because a) chemically they are typically indistinguishable and b) neither $Gd^{3+}$ nor $Tb^{3+}$ are reduced nor oxidized at reasonable voltages. Periodic Boundary Conditions (PBC) replicate in the XY plane the structure that has been explicitly defined in the main calculation box. This allows only a partial simulation of the intermolecular interactions, which is limited by the artificially imposed rectangular packing and by the XY size of the junction box.

The convergence of the electronic structure of the leads is achieved with 2x2x128 Monkhorst-Pack k-point mesh, while for the SR one sets open boundary conditions in the transport direction and periodic ones along the transverse plane, for which an identical k-point mesh is used (2x2x1 k-points). The exchange-correlation potential is described by the GGA (PBE) functional. The Au-valence electrons are represented over a numerical s-only single-$\zeta$ basis set that has been previously demonstrated to offer a good description of the energy region around the Fermi level. In contrast, for the other atoms (C, O, N, H, and Gd), we use a full-valence double-$\zeta$ basis set. Norm-conserving Troullier-Martins pseudopotentials are employed to describe the core-electrons in all cases. Finally, the spin-dependent current, $I_\sigma$, flowing through the junction is calculated from the Landauer-Büttiker formula,

$$I_\sigma(V) = \frac{e}{h}\int_{-\infty}^{+\infty} T_\sigma(E,V)[f_L(E-\mu_L) - f_R(E-\mu_R)]dE$$

where the total current $I_{tot}$ is the sum of both the spin-polarized components, $I_\sigma$ being $\sigma = spin\ up/spin\ down$. Here $T_\sigma(E,V)$ is the transmission coefficient and $f_{L/R}$ are the Fermi functions associated with the two electrodes chemical potentials, $\mu_{L/R} = \mu_o \pm V/2$, where $\mu_o$ is the electrode common Fermi level.

Additional transport calculations considering an applied bias voltage V = 0.5 V and 1.0 V (see DoS and transmission spectrum in Fig. S11.1) show the typical displacement of the valence bands to lower energies, meaning that, in the absence of a gate voltage, higher bias voltages are needed to achieve favorable conduction channels. The current obtained for such conditions with the LB formula are, $0.069 \cdot 10^{-6}$ mA and $0.35 \cdot 10^{-6}$ mA respectively. These values are grossly overestimating the total current intensity. This is a common occurrence in NEGF calculations and is usually attributed to the absence of an adequate exchange-correlation functional that could account for accurate single particle levels using Kohn-Sham orbitals.[9]

Finally, let us comment on the participation of the magnetic orbitals of the lanthanide ion in the conduction path near Fermi Energy. In principle with this result, it is possible to probe the implication of the electrode orbitals on the conduction channels. (see Fig. S11.2) This, with the DOS analysis, and considering the width of the peaks as an indication of the interaction, makes it possible to conclude that the hybridization between Au and the GdLBTC biomolecule is very low. This behavior coincides with the previous calculations on thiol-based attachments performed by some of the authors.[10]

---

[9] Ullrich, C. A. Time-Dependent Density-Functional Theory: Concepts and Applications, **2012**, *Oxford University Press*.

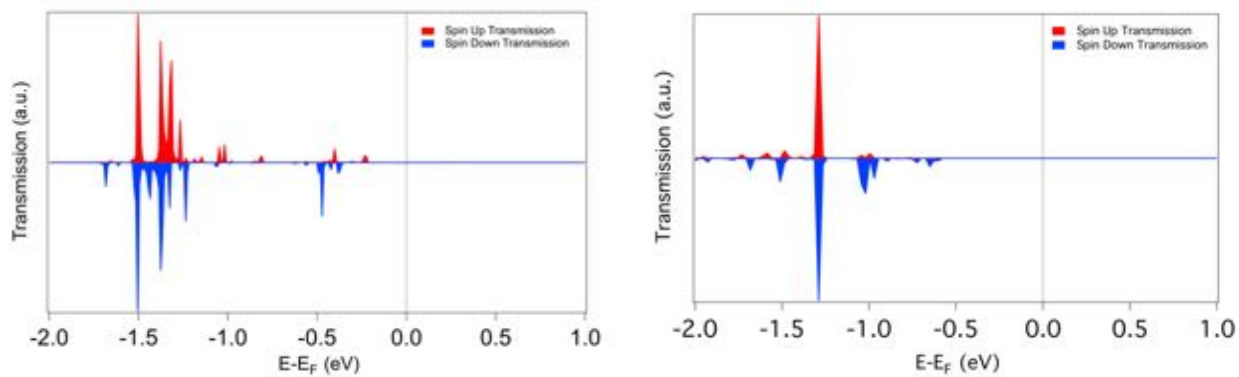

**Figure S11.1**. Transmission spectra of the total scattering region. (left): calculation with bias voltage 0.5 V. (right), calculation with bias voltage 1.0 V.

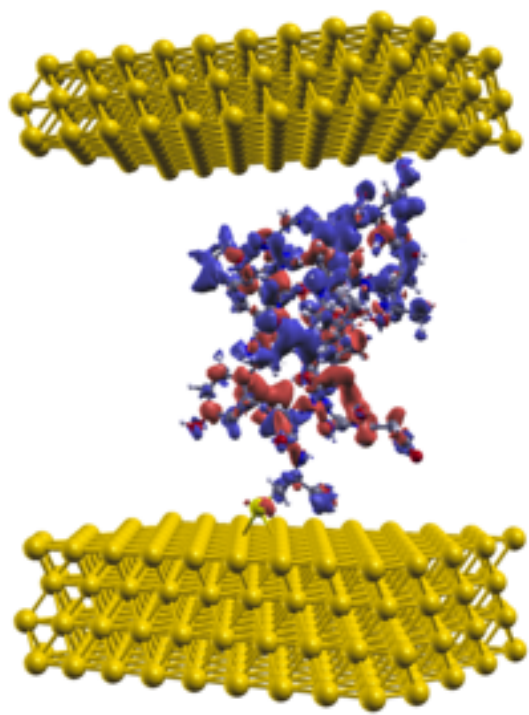

**Figure S11.2**. Local Density of States at the main conducting peak energy, serving as a visual guideline of the electron path. Most of the density is located in the π orbitals of the peptide amide bonds, therefore following the roughly helicoidal shape of peptide.

---

[10] Cardona-Serra, S.; Gaita-Ariño, A.; Stamenova, M.; Sanvito, S. Theoretical Evaluation of [$V^{IV}(\alpha\text{-}C_3S_5)_3$]$^{2-}$ as Nuclear-Spin-Sensitive Single-Molecule Spin Transistor *J. Phys. Chem. Lett.* **2017**, *8* (13), 3056–3060.

# S12 Determination of the magnetic interaction between the spin polarized current and the spin-orbit momentum of $Tb^{3+}$

In order to understand the magnetic interaction between the spin current and the, in principle, paramagnetic anisotropic $Tb^{3+}$ spin-orbit momentum, one should first consider a situation which will be explained between two opposite scenarios. In the first, we will consider a fixed momentum on the $Tb^{3+}$, this means a strong preferential orientation and thus a noticeable filtering effect on the spin current. In the second, we will consider the lanthanoid being in completely thermal equilibrium, thus meaning that all the magnetic levels split by anisotropy are quasi equally populated and the system behaves as a paramagnet with no preferential orientation in absence of external magnetic field. In our experimental setup, the effect is seen to occur between these two extreme scenarios.

In a first step we estimated the spin filtering effect created by a fully polarized lanthanide ion on the current. In this case, considering the small peak at $E-E_F$ = -0.3 eV in the transmission spectra (see Figure S12.1) and, assuming a complete polarization of the lanthanide ion, a value of SP = -70%. This will only be realistic as extreme scenario when assuming that the lanthanide ion is completely polarized by an external magnetic field, which in fact is distant from the realistic conditions in our experiments. This value should be considered as an upper limit of the spin filtering effect produced by the lanthanoid ion.

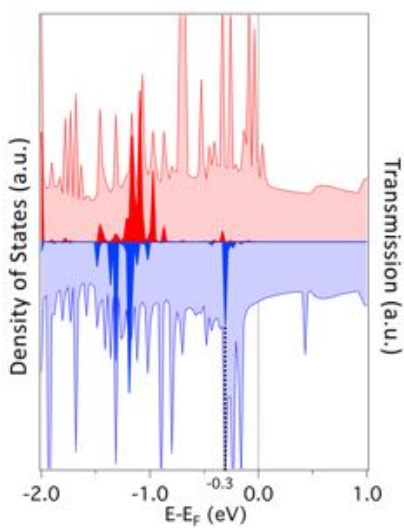

**Figure S12.1**. Calculated Density of States (light colors) and normalized transmission spectra (strong colors) at zero voltage, distinguishing between spin-up (red) and spin-down (blue), where the up-down reference is the polarization direction of the f electrons in the lanthanide. A sharp spin-down conduction peak near the Fermi level is marked with dashed line as the main conducting peak at low gate voltages.

In a completely opposite situation, we estimate the magnetic interaction between the spin current and the lanthanoid ion in thermal equilibrium. This calculation gives a more realistic insight on the magnitude of the magnetic interaction between the electronic current passing through the junction and the magnetic momentum of the lanthanoid ion in an ideal steady state in equilibrium which is neither our experimental case. To perform this calculation we first-principles DFT calculations using the Broken Symmetry approach. The LBTC structure in the most stable configuration between the Au electrodes was used without further modification. For the sake of simplicity and, as we expect that magnetic interaction will not be strongly modified by the electrode-molecule interaction, we performed the calculations without considering the electrodes explicitly. For the estimation we used the GAUSSIAN16 code, a well-known tool for molecular DFT calculations.[11] With respect to the spin-orbit coupling we used Douglas-Kroll-Hess $2^{nd}$ order scalar relativistic methodology[12] with a UB3LYP exchange-correlation functional and a basis set consisting in 6-31G** for C, H, N and O and SARC-DKH2[13] for $Tb^{3+}$.[14] We proposed an initial orbital guess where the extra electron is mainly located in the amino acids that directly coordinate the lanthanoid. This would represent the electronic spin current flowing through the device with its magnetic spin either parallel or antiparallel to the $Tb^{3+}$ spin-orbit momentum. The difference in total energies obtained comparing these two calculations is: **4.32cm$^{-1}$**. The antiparallel alignment between the lanthanide spin and the spin on the peptide is the most stable according to this calculation, in complete agreement with the predicted filtering in LBTC where the transport channel nearest to $E_F$ has a reverse polarization compared with the momentum of the lanthanoid. Considering this energy level splitting between the Ferro and Antiferromagnetic alignment, we obtained a bottom limit of SP' = -0.1%, experimentally indistinguishable from SP in absence of a paramagnetic ion. This result obviously lacks some important characteristics of the process that occur during the experimental observation which in fact is not a steady state at all. However, it could serve as a lower estimation of the minimum filtering effect one could expect.

---

[11] Gaussian 16, Revision C.01, Frisch, M. J.; Trucks, G. W.; Schlegel, H. B.; Scuseria, G. E.; Robb, M. A.; Cheeseman, J. R.; Scalmani, G.; Barone, V.; Petersson, G. A.; Nakatsuji, H.; Li, X.; Caricato, M.; Marenich, A. V.; Bloino, J.; Janesko, B. G.; Gomperts, R.; Mennucci, B.; Hratchian, H. P.; Ortiz, J.

To explain and understand our experimental observation, where a net effect of the lanthanoid on the spin polarization is evident, one needs to realize that the experimental reality lies between the previous two scenarios. Even considering a modest spin polarized current of 1nA passing through a single molecule would mean approximately $6\cdot10^9$ spin carriers interacting antiferromagnetically with the lanthanoid momentum per second. This widely surpasses the spin relaxation rate of the lanthanoid (~ns) and will naturally contribute to maintain the orientation of the lanthanoid on a fixed out-of-equilibrium orientation. This effect mimics the well-known Spin-torque effect mostly present in all current spintronic effects like TMR and GMR. This non-equilibrium situation would imply that the relaxation via thermalization of the magnetic momentum of the $Tb^{3+}$ ion is not possible and that the spin current contributes to enhance the polarization of the Tb spin, thus contributing to a larger observable effect in terms of spin filtering.

V.; Izmaylov, A. F.; Sonnenberg, J. L.; Williams-Young, D.; Ding, F.; Lipparini, F.; Egidi, F.; Goings, J.; Peng, B.; Petrone, A.; Henderson, T.; Ranasinghe, D.; Zakrzewski, V. G.; Gao, J.; Rega, N.; Zheng, G.; Liang, W.; Hada, M.; Ehara, M.; Toyota, K.; Fukuda, R.; Hasegawa, J.; Ishida, M.; Nakajima, T.; Honda, Y.; Kitao, O.; Nakai, H.; Vreven, T.; Throssell, K.; Montgomery, J. A., Jr.; Peralta, J. E.; Ogliaro, F.; Bearpark, M. J.; Heyd, J. J.; Brothers, E. N.; Kudin, K. N.; Staroverov, V. N.; Keith, T. A.; Kobayashi, R.; Normand, J.; Raghavachari, K.; Rendell, A. P.; Burant, J. C.; Iyengar, S. S.; Tomasi, J.; Cossi, M.; Millam, J. M.; Klene, M.; Adamo, C.; Cammi, R.; Ochterski, J. W.; Martin, R. L.; Morokuma, K.; Farkas, O.; Foresman, J. B.; Fox, D. J. Gaussian, Inc., Wallingford CT, **2016**.
[12] Douglas, M.; Kroll, N. M.; Quantum electrodynamical corrections to fine-structure of helium, *Ann. Phys.* **1974**, *82*, 89-155.
[13] Pantazis, D. A.; Neese, F.; All-Electron Scalar Relativistic Basis Sets for the Lanthanides. *J. Chem. Theory Comput.*, **2009**, *5*, 2229-2238.
[14] (a) Becke A. D.; A new mixing of Hartree-Fock and local density-functional theories. *J. Chem. Phys.* **1993**, *98*, 1372–1377. (b) Lee, C.; Yang, W.; Parr, R. G. Development of the Colle-Salvetti correlation-energy formula into a functional of the electron density. *Phys. Rev. B*. **1988**, *37*, 785–789.